\documentclass[]{pasj01}
\Received{}
\Accepted{2024/06/25}
 
\usepackage[switch,mathlines]{lineno}
\usepackage{longfigure}

\usepackage{dcolumn}

\definecolor{bv}{rgb}{0.54, 0.17, 0.89} 
\definecolor{hg}{rgb}{0.0, 0.44, 0.0} 

\begin{document} 
\title{
Spiral Magnetic Field and Their Role on Accretion Dynamics in the Circumnuclear Disk of Sagittarius A*: Insight from $\lambda=$ 850 $\mu$m Polarization Imaging}

\author{Kazuki \textsc{Sato}\altaffilmark{1}}
\altaffiltext{1}{Physics and Astronomy Program, Graduate School of Science and Technology, Kagoshima University, 1-21-35 Korimoto Kagoshima, Kagoshima, 890-0065,Japan}
\email{kadai6961489@gmail.com, 
shinnaga@sci.kagoshima-u.ac.jp, rsf@tokushima-u.ac.jp, stakeru@ea.c.u-tokyo.ac.jp, kakiuchi@g.ecc.u-tokyo.ac.jp, jott@nrao.edu}
\author{Hiroko \textsc{Shinnaga}\altaffilmark{1,2} \thanks{Corresponding author.}}

\altaffiltext{2}{Amanogawa Galaxy Research Astronomy Center (AGARC), Graduate School of Science and Technology, Kagoshima University, 1-21-35 Korimoto Kagoshima, Kagoshima, 890-0065,Japan}
\author{Ray \textsc{S. Furuya}\altaffilmark{3,4}}
\altaffiltext{3}{Institute of Liberal Arts and Sciences, Tokushima University, Minami Jousanajima-machi 1-1, Tokushima, Tokushima, 770-8502, Japan}
\altaffiltext{4}{Visiting Associate Professor, National Astronomical Observatory of Japan, Mitaka, Tokyo, Japan}
\author{Takeru \textsc{K. Suzuki}\altaffilmark{5}}
\altaffiltext{5}{School of Arts and Sciences, University of Tokyo, 3-8-1 Komaba, Meguro, Tokyo 153-8902, Japan}
\author{Kensuke \textsc{Kakiuchi}\altaffilmark{5}}
\author{Jürgen \textsc{Ott}\altaffilmark{6}}
\altaffiltext{6}{National Radio Astronomy Observatory, 1003 Lopezville Road, Socorro, NM 87801, U.S.A.}

\maketitle
\begin{abstract} 
We showcase a study on the physical properties 
of the Circumnuclear Disk (CND) surrounding the SMBH Sgr A* of 
the Galactic Center, emphasizing the role of magnetic field  ($\vec{B}$ field) with 0.47 pc spatial resolution, 
based on the sensitive $\lambda=$ 850 $\mu$m polarization data 
taken with the JCMT 
SCUBA2/POL2. 
We analyzed 
ancillary 
datasets: CS $J=2-1$ emission taken with the 
ALMA, 
continuum emissions taken at $\lambda$ = 6 cm 
and at $\lambda$ = 37 $\mu$m taken with the VLA and SOFIA. 
The $\vec{B}$ field within the CND exhibits a coherent spiral pattern. 
Applying the model described by Wardle and K\"{o}nigl 1990 (WK model) to the observed $\vec{B}$ field pattern, 
it favors gas-pressure-dominant models without dismissing a gas-and-$\vec{B}$ field comparable model, leading us to estimate the $\vec{B}$-field strength in the ionized cavity around Sgr A* as $0.24^{+0.05}_{-0.04}$ mG. 
Analysis based on the WK model further allows us to derive representative $\vec{B}$-field strengths for the radial, azimuthal, and vertical components as $(B_r, B_\phi, B_z) = (0.4 \pm 0.1, -0.7 \pm 0.2, 0.2 \pm 0.05)$ mG, respectively. 
A key finding is that the $|B_\phi|$ component is dominant over $B_r$ and $B_z$ components, consistent with the spiral morphology, indicating that the CND's $\vec{B}$-field is predominantly toroidal, possibly shaped by accretion dynamics. 
Considering the turbulent pressure, estimated plasma $\beta$ values indicate the effective gas pressure should surpass the magnetic pressure. 
{Assessing the CND of our MWG in the toroidal-and-vertical stability parameter space, we propose that such an “effective” magnetoro-tational instability (MRI) may likely be active.}
The estimated maximum unstable wavelength, $\lambda_{\rm max} = 0.1 \pm 0.1$ pc, is smaller than the CND's scale height ($0.2 \pm 0.1$ pc), which indicates the potential for the effective MRI intermittent cycles of $\sim$ $10^6$ years, %
which should 
profoundly affect the CND's evolution, considering the 
estimated mass accretion rate of 
$10^{-2} M_{\odot}$ yr$^{-1}$ to the SMBH.
\end{abstract}

\section{Introduction}
\label{s:Intro}
\subsection{SMBH and $\vec{B}$ Field}
\label{ss:Intro_BH}

SMBHs at galaxies' centers hold masses ranging from 0.1\% to 1\% of their host galaxies, and their evolution is intricately linked with that of the galaxies themselves (e.g., \cite{Ding2020, Mountrichas2023} and references therein). 
The 
{mass increase} of SMBHs is primarily due to gas accretion onto structures known as active galactic nuclei (AGNs) and through mergers between SMBHs that accompany galaxy mergers. 
The former scenario is particularly intriguing regarding the cause and physics behind mass accumulation. 
Given the typical mass accretion rate of an order of 1 \(M_{\odot}\) yr\(^{-1}\) to SMBHs in AGNs \citep{Kormendy2013}, the accretion phase in their evolution would be an order of \(10^8\) years, which is remarkably short compared to the age of the universe \citep{Yu2002, Shirakata2019}.
Indeed, only about 1\% of observed galaxies are known to harbor AGNs  high-luminosity quasars (e.g., \cite{Alexander2012, Aird2015}), aligning with this accretion mechanism.
Moreover, because almost all galaxies harbor SMBHs at their centers, mass accretion may be an episodic-multiple processes with respect to their lifetime, with periods of both under-and-over accumulations (e.g., \cite{Li2023}). 
Identifying what triggers or prevents mass accretion is a crucial question to understanding the physics within the accretion disk of an AGN and the transport of matter towards the Galactic Center.

A typical set of structures and phenomena often observed in the vicinity of SMBHs include accretion disks, disk winds, and jets, all of which constitute the black hole's magnetosphere. 
To understand this system, it is essential to comprehend the velocity field and angular momentum transport within the accretion disk \citep{Lubow1994}, similar to the situation in disks observed around young stellar objects (e.g., \cite{Okuzumi2014}).
Notably, the amplification of the $\vec{B}$ field inside the accretion disk, driven by differential rotation and followed by cycles of the $\vec{B}$ field's saturation and decay, determines the physical properties of the accretion flow.
Theoretically, magnetic rotational instability (MRI) has been proposed \citep{Vel59, cha61, BH91}, leading to exponential $\vec{B}$-field amplification within a few rotations in the presence of a seed $\vec{B}$ field, effectively converting kinetic energy into magnetic energy. 
The growth rate of the instability varies with wavelength and the wavelength of the mode that provides the fastest maximum growth rate ($\lambda_{\rm max}$) is given by the Alfv\'en velocity divided by the rotational-angular velocity. 
Observationally, verifying this requires spatial resolution capable of resolving the maximum growth wavelength. 
Such an amplification of $\vec{B}$-field causes angular momentum transport within the disk, and how the excess-angular momentum is transported outward dictates the accretion rate and $\vec{B}$-field's strength.
During the $\vec{B}$-field amplification, the outward transport of angular momentum causes gas that has lost angular momentum to fall inward, determining the mass accretion rate. 
Once the energy of the amplified $\vec{B}$ field equals the thermal energy, the $\vec{B}$-field saturates and decays through dissipation via magnetic reconnections, with the remaining $\vec{B}$ field serving as a new seed for the cycle (e.g., \cite{SB1996, Pessah2007, Sano2004, Latter2015, Kudoh2020}).

Recent theoretical interest has focused on Magnetically Arrested Disks (MADs) proposed by \citet{Narayan2003} where radially supported by a rotating axial component of the $\vec{B}$ field accumulated through mass accretion.
When such a $\vec{B}$-field threatens a rotating black hole, the rotational energy of the black hole is extracted along the $\vec{B}$ field lines, ejecting a relativistic jet (e.g., \cite{Igumenshchev2008, Tchekhovskoy2011, KS2018}). 
This process, known as the Blandford-Znajeck mechanism \citep{BZ1977}, is a promising model for driving relativistic jets. 
Therefore, unveiling the  $\vec{B}$-field structure and estimating their strength supplied from the accretion disk to the vicinity of the black hole is crucial for grasping the driving mechanisms of the black hole magnetosphere and relativistic jets. 
As the accretion rate onto the black hole increases, radiative cooling becomes significant, and the cooling may eventually surpass the gravitational energy release rate from accretion. 
In such disks, the azimuthal component of the $\vec{B}$ field is predicted to be amplified, leading to a disk supported by the azimuthal $\vec{B}$ field, termed a ``toroidal MAD", compared to a MAD supported by a vertical $\vec{B}$ field \citep{MamiMachida2006}. 
Regardless of the achievable spatial resolution, knowledge of the $\vec{B}$ field structure of the black hole magnetosphere is essential for understanding accretion and jet-driving mechanisms, posing significant constraints on theoretical models (e.g., \cite{Roche2018, GRAVITY 2020, EHT2021M87Bfield, Gravity2023SgrA}).

\subsection{The Galactic Center $\vec{B}$ Field and Sagittarius A*}
\label{ss:Intro_BH}

Here we present our study on the gas dynamics surrounding Sagittarius A* (Sgr A*; \cite{bal74}), the nearest SMBH candidate (e.g., \cite{eck97, ghe98, ghe03, gen10}). 
The $\vec{B}$-field structure at the center of the Milky Way Galaxy (MWG) 
will allow us to estimate 
$\vec{B}$-field strength.  
Understading the $\vec{B}$-field of the Galactic Center is crucial, 
in terms of the evolution of the interstellar medium (ISM) in the Galactic disk \citep{Inutsuka2015}.
It is generally known that spiral galaxies have a 
{global $\vec{B}$-field} structure along their spiral arms, as evidenced by polarimetric imaging of the centimeter (cm) wavelength Synchrotron emission \citep{Beck2015}. 
Based on observations such as pulsar polarization, the $\vec{B}$-field strength of such disk galaxies is typically estimated to be a few $\mu$G. 
Rotation Measures (RMs) from polarization observations at multi-wavelength cm imaging allow to plot RMs as a function of azimuthal angles around the galactic center, enabling the 
{global $\vec{B}$-field} structure in spiral galaxies to be either Axis Symmetric Spiral (ASS) or Bi-Symmetric Spiral (BSS)\citep{TS78, Sofue1986}.
The reasons for this classification are not fully understood but can be broadly categorized into theories of primordial field origin at the time of galaxy formation and galaxy dynamo theory \citep{Parker1955, Brandenburg2023}. 
If a galaxy's primordial $\vec{B}$ field were perpendicular to the galactic plane of the protogalaxy, the ASS type would likely form. 
At the same time, a parallel orientation would favor the formation of the BSS type \citep{Sofue2010}.
Another explanation is based on the galaxy-dynamo theory, which posits that the $\vec{B}$-field was formed simply according to the induction equation of Electromagnetism.
Generally, the axially symmetric component of a galaxy's internal motion tends to form an ASS-type field. 
In contrast, the BSS-type field tends to form when turbulent motion or the off-axis component of gas dominates. 
The Milky Way's $\vec{B}$ field is classified as BSS-type, but its origins and growth process remain largely unknown (e.g., \cite{Widrow2002, Brandenburg2023}). 
The unresolved situation regarding the origins and growth process of the galaxy-wide $\vec{B}$ field is not unique to the MWG 
(\cite{Garaldi2021}).

Given the above, deepening our understanding of the $\vec{B}$-field structure at the center of the MWG is fundamentally important for advancing our knowledge of spiral galaxies in general and the evolution of the ISM within these galaxies. 
This remains an open question across various scientific fields. 
At the Galactic Center, a Kerr black hole with a mass of $4.152\times 10^6 M_\odot$ (\cite{EHT2022}; see also \cite{Do2019, GRAVITY2022}) is located at a distance of $8178\pm 13_{\rm stat}\pm 22_{\rm sys}$ pc \citep{GRAVITY2019}. 
The compact radio source, Sgr A*, embeds the SMBH, generating strong tidal forces and $\vec{B}$ field. 
Sgr A* is located near the center of a minispiral structure, comprising ionized gas and dust streams \citep{eke83, KYLo83, Zhao2009, Zhao2010, Tsuboi2016}, and the circumnuclear disk (CND), a dense aggregation of gas and dust \citep{chr05, mar12, tsu18}.
Next, we provide a brief overview of the CND and the gas in its vicinity before delving into studies of its $\vec{B}$ field.

The CND, a parsec-scale ring-like structure \citep{jac93, lau13}, orbits at a rotational velocity of approximately $100$ km s$^{-1}$ around Sgr A*. 
A portion of the surrounding gas is believed to be infalling from the CND to Sgr A* \citep{mon09}. 
The neutral gas constituting the CND is characterized by high turbulence, density, and temperature \citep{Herrnstein2002, Mills2013}. 
The CND gas also exhibits azimuthal and radial abundance changes of molecules \citep{mar12}. 
\citet{Hsieh2017} conducted 0.05 pc multiple-transition CS line imaging with ALMA and revealed that the CS molecular cloudlets range in size from 0.05 to 0.2 pc, with a wide-velocity-dispersion range of $5\lesssim \sigma/[{\rm km s^{-1}}] \lesssim 40$. 
They found the compact CS molecular clouds comprise high-density warm components (kinetic temperature $50\lesssim T_{\rm k}/[{\rm K}] \lesssim 500$) and very-high-density cold components ($T_{\rm k} \lesssim 50$ K). 
Here, ``high-density" refers to $10^3 \lesssim n_{\rm H_2}/[{\rm cm}^{-3}] \lesssim 10^5$, and ``very-high-density" to $10^6 \lesssim n_{\rm H_2}/[{\rm cm}^{-3}] \lesssim 10^8$. 
It should be noted that the thermal emission of dust grains at 850 $\mu$m wavelength, which is used in our magnetic-field study reported in this paper, traces the ``high-density" but cold gas, diverging from the classifications provided by \citet{Hsieh2017}.
\citet{Hsieh2017} also presented a stability analysis of the CS-emitting gas, claiming that the majority (${84}_{-37}^{+16}\%$) of the CND gas, located further than $\approx 1.5$ pc from Sgr A*, is tidally stable due to turbulence support.
Earlier observations with a lower resolution before the ALMA era indicated the presence of streaming gas to the CND from surrounding structures traced by NH$_3$ lines \citep{Coil2000, Minh2013_NH3streaming}. 
Shock-enhanced SiO emission was detected west of Sgr A West, potentially where several incoming and outgoing gas flows interact \citep{Minh2015_SiO}. 
This interpretation is supported by \citet{Hsieh2017}, who indicated that a gas-feeding mechanism to the CND is active within a radius range of 20 pc to 2 pc centered on Sgr A*.
Estimating gas properties in the Galactic Center region is fundamentally challenging due to the 8 kpc distance and line-of-sight contamination. 
We must always be mindful of such caveats and the possibility that subsequent observations may revise our knowledge.

Utilizing the NASA Kuiper Airborne Observatory, \citet{hil90} presented a far-infrared (FIR) 100 $\mu$m polarization map of thermal emission from dust grains at six positions within the dust ring of Sgr A. 
The $\vec{B}$ field angles, predominantly perpendicular to the ring's long axis, suggest that the $\vec{B}$ field lies primarly within the ring's plane, resembling the pattern expected for a magnetized accretion disk. 
\citet{hil90} pointed out that centrifugal acceleration-driven energy and angular momentum may be removed along $\vec{B}$ field lines.
Based on the observations by \citet{hil90}, \citet{war90} developed a self-similar model of a \(\vec{B}\) field in the molecular disk, initially presented in \cite{kon89}, to predict the polarization properties of dust emission in the far-infrared and the Zeeman-effect absorption-line profiles. 
The modeling by \citet{war90} revealed that the radial ($B_r$) and azimuthal ($B_\phi$) $\vec{B}$ field components within the CND are comparable in magnitude but have opposite signs, indicative of $B_\phi$ generation from $B_r$ through shear motion. 
\citet{war90} also found a minor vertical component ($B_z$) compared to $B_r$ and $B_\phi$, facilitating angular momentum removal via centrifugally driven outflow.
Furthering the FIR researches as well as \citet{Hildebrand1993, Aitken2000}, \citet{Hsieh2019} analyzed 850 $\mu$m SCUPOL dust polarization data from the James Clark Maxwell Telescope (JCMT) to study the $\vec{B}$-field structure within the CND. 
Despite SCUPOL being an earlier generation polarimeter than those used in our study, their analysis corroborated the 100 $\mu$m polarimetric findings and aligned with 350 $\mu$m observations by \citet{Novak2000}. 
Employing the \citet{war90} model, \citet{Hsieh2019} presented a coherent $\vec{B}$-field structure across the CND. 
With a line-of-sight $\vec{B}$-field strength estimation of 1mG from HI Zeeman effect measurements by \citet{Plante1995}, they proposed a plasma $\beta$ of $\lesssim 1$, stressing the $\vec{B}$-field's pivotal role in the CND's gas dynamics and its influence extending to the inner mini-spiral. 
Here, \(\beta\) is defined as the ratio of gas pressure \(P_{\rm gas}\) to the magnetic pressure, with \(\beta \equiv \frac{P_{\rm gas}}{|\vec{B}|^2/(8\pi)}\) in cgs units.
Recently, \citet{gue23} refined the plane-of-sky $\vec{B}$-field strength estimates using 53 $\mu$m polarimetric continuum images at 4\farcs85 resolution, about 0.18 pc, via SOFIA/HAWK+. 
Their advanced Davis–Chandrasekhar–Fermi (DCF) method, incorporating large-scale shear flow effects, identified $\vec{B}$-field strengths between 1-27 mG and discussed the shift from magnetic to gravitational dominance in material accretion onto Sgr A* starting at approximately 1 pc.

\subsection{Structure of This Paper}
\label{ss:Intro_Structure}

This paper is structured as follows: \S\ref{s:Data} details the data retrieval and reduction processes for the JCMT polarization data and the ALMA molecular line observations. 
\S\ref{s:ResAna} presents our results, including the $\vec{B}$ field map of the CND, its comparison with the \citet{war90} (WK) model, and the molecular gas analysis. 
\S\ref{s:D} discusses our findings' implications for the $\vec{B}$ field's role in the CND, especially regarding magnetorotational instability (MRI) as a possible mechanism of angular momentum transport. 
We conclude with a summary in \S\ref{s:Summary}. 

\section{Data Retrieval and Reduction}
\label{s:Data}
Given the scientific goals, we generated polarization images using two archival data sets acquired from the James Clerk Maxwell Telescope (JCMT; 
\S\ref{ss:pol2data}).
To interpret the data from JCMT, we utilized molecular line data from the archive of the Atacama Large Millimeter/submillimeter Array (ALMA; \S\ref{ss:ALMA}) and continuum emission data from the archives of the Very Large Array (VLA) and the Stratospheric Observatory for Infrared Astronomy (SOFIA; \S\ref{ss:VLAandSOFIAdata}).
The key parameters that clarify the roles of these data sets are summarized in Table \ref{tbl:Obs}.

\subsection{The $\lambda=$ 850 $\mu$m Polarization Data from the JCMT}
\label{ss:pol2data}

The archival data were initially obtained by P. T. P. Ho (Proposal ID: M16BP061) and G. Bower (M17AP074) over the course of nine nights, consisting of 29 exposures, under the conditions of JCMT's Weather Band 2.
The $850\,\mu {\rm m}$ polarization data were collected using the POL-2 polarimeter, which was mounted in front of the SCUBA-2 \citep{hol13} on the JCMT.

Eight out of twenty minimum science blocks (MSBs) were carried out in August 2016, and the remaining MSBs were obtained in March 2017, yielding a total integration time of 14.6 hours.
All the MSBs, except for one, were performed under the conditions of JCMT's Weather Bands 1 or 2 (zenith opacity at 183 GHz, $\tau_{183\mathrm{GHz}}\leq$ 0.08); the exception was one MSB taken under $\tau_{183\mathrm{GHz}} =$ 0.084. All observations were conducted within an airmass range of 1.51 to 1.94.
To produce Stokes $I, Q,$ and $U$ images in the equatorial coordinate system, we reduced the 850 $\mu$m time series data using the \texttt{pol2map.py} script, which is implemented in the SMURF package of the Starlink suite \citep{cha13}. We followed the standard pipeline (version dated July 23, 2021) as described in, e.g., \citet{war17}.
Instrumental polarizations were subtracted using the 2019 model.

In this paper, we adopt the updated beam size \citep{mai21} of 12\farcs{6} in HPBW, which corresponds to 0.50\,pc at a distance $d$ of 8.178\,kpc.
Given that our target is more extended than the beam size, we applied the updated flux-conversion factor of \(2.19\pm 0.13\) Jy arcsec\(^{-2}\) pW\(^{-1}\) \citep{mai21} to ensure that the resultant images are scaled to the specific intensity.
The number used is the geometrical mean of the ``Aperture FCFs'' for the corresponding observing periods listed in Table 4 of \citet{mai21}.
Furthermore, we applied an attenuation factor of \(1.35\pm0.008\) (P. Friberg, private comm.) to account for the insertion of the POL-2 system in front of the focal plane of SCUBA-2.
We debiased the positive noise in polarized intensity, \(PI\), using a modified asymptotic estimator as described by \citep{pla14,mon15}.
We estimate that the resultant Stokes \(I\) map has achieved a noise level of 15.8 mJy arcsec\(^{-2}\) with a 4\farcs0-pixel size over the central region of approximately \( \sim 3\farcm0 \) in diameter; the noise levels for the Stokes \(Q\) and \(U\) maps were 15.5 mJy arcsec\(^{-2}\) and 19.1 mJy arcsec\(^{-2}\), respectively.
After completing all the processes, we converted the data from equatorial coordinates to Galactic coordinates to facilitate comparison with previous works.

\subsection{The CS $J=2-1$ line data from the ALMA}
\label{ss:ALMA}
We retrieved the archival ALMA data taken by \citet{tsu18} (Table \ref{tbl:Obs}) and calibrated the visibility data using the Common Astronomy Software Applications (CASA v5.1.1-5) pipeline along with calibration scripts for Cycle 5 data.
To combine the visibility data from both the main 12\,m-dish array and the 7\,m-dish Atacama Compact Array (ACA), we used the 'concat' task in CASA v6.4, followed by the subtraction of continuum emission.
Image construction and deconvolution using interferometric visibilities, as well as single-dish ones from the total power array, were carried out using the 'tclean' and 'feather' tasks in CASA (v6.4) with a Briggs robust parameter of 0.5.
The automasking function in 'tclean' was utilized during the cleaning process.
The final image cubes, which have units of specific intensity ($I_{\nu}$), were created with a velocity resolution of 2.0 km s$^{-1}$.
Finally, we converted the units of $I_{\nu}$ to the beam-averaged brightness temperature $T_{\rm b}$, measured in Kelvin, using the formula $T_{\rm b} = \frac{I_{\nu} \lambda^{2}}{2k_{\rm B}}$, where $k_{\rm B}$ denotes the Boltzmann constant.
This conversion resulted in a value of 28.55 K sr$^{-1}$ for the 2\farcs67$\times$1\farcs67 beam at a frequency of $\nu = 97.980$ GHz

In \citet{tsu18}, the authors presented results from multiple molecular line analyses using ALMA's 12-m and 7-m arrays (see Table \ref{tbl:Obs} for the corresponding project codes).
As the focus of this paper is on the results from the polarimetric observations at JCMT, we limited our analysis to the sole CS $J=2-1$ transition.
Lastly, we emphasize that our enhanced image cube, produced from the 12\,m, 7\,m, and total power (TP) arrays, is overall consistent with the one used in \citet{tsu18}, which did not include TP data.
The resultant RMS image noise level is described in the caption of Figure \ref{fig:CS21on850and37}.

\subsection{The $\lambda=$ 6\,cm Continuum Emission Data from VLA and $\lambda=$ 37.1\,$\mu$m Continuum Emission Data from SOFIA}
\label{ss:VLAandSOFIAdata}
From the NRAO data archive, we retrieved the already-reduced image taken in the $\lambda = 6$ cm band with the C-array configuration over a 2-hour period on April 3, 1992 (see Table \ref{tbl:Obs}).
The $\lambda=$ 37 $\mu$m imaging at SOFIA allows for a high dynamic range in brightness, which is essential for detecting faint emission \citep{deb17}.  Thus, we selected the $\lambda=$ 37 $\mu$m band and retrieved the fully-reduced $\lambda = 37.1 \, \mu$m continuum emission image from the SOFIA data archive. 
The data were originally acquired during the seventh cycle as part of the FORCAST Galactic Center Legacy Project \citep{han20} (see Table \ref{tbl:Obs}).

\section{Results and Analysis}
\label{s:ResAna}

In this section, we present the data and describe the results based on our analysis on the $\lambda=$ 6\,cm and 37.1\,$\mu$m continuum images, as well as the $\lambda=$  850\,$\mu$m continuum data with polarimetry. 
We revealed that the 37.1\,$\mu$m continuum emissions represent thermal radiation from dust grains surrounding Sgr\,A* and the $\lambda=$ 6\,cm emission traces components of synchrotron radiation originating from the SMBH activity, peaked at the position of the SMBH,  along with dust components, while the $\lambda=$ 850\,$\mu$m continuum emission pertains to the body of the CND.

\subsection{Comparisons Between the $\lambda=$ 6\,cm and $\lambda=$ 850\,$\mu$m Continuum Maps and Between the  $\lambda=$  37.1\,$\mu$m and $\lambda=$ 850\,$\mu$m Maps}
\label{sss:ContImages}

\subsubsection{Morphological Comparisons}
\label{sss:MorComp}
\begin{figure}[ht]
\begin{center}
\includegraphics[width=8.2cm]{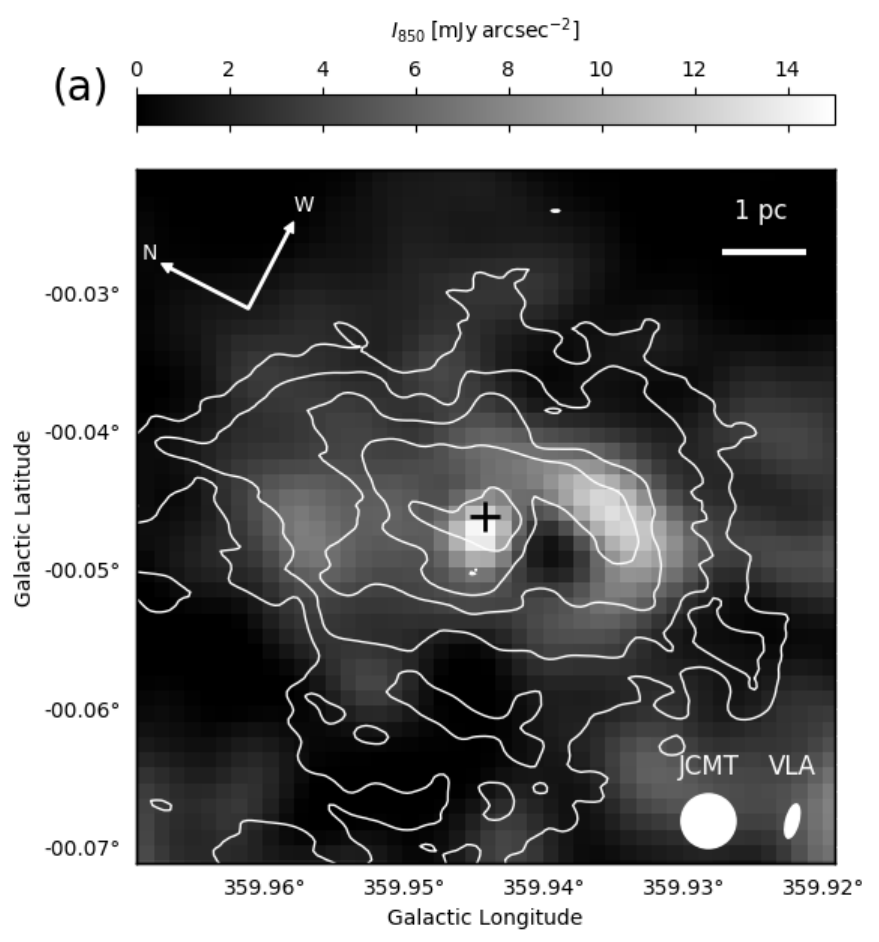}
\includegraphics[width=8.2cm]{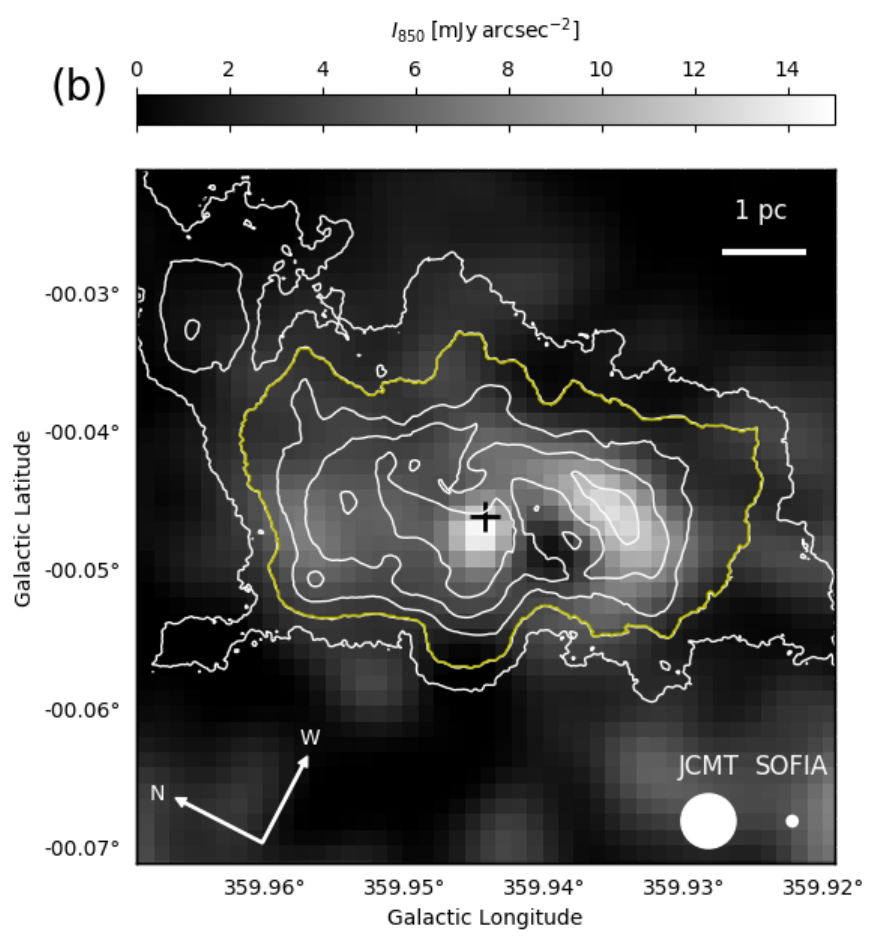}
\end{center}
\caption{Overlays of (a) the $\lambda =\,6$ cm continuum (contours) and (b) $\lambda=$ 37.1 $\mu$m (contours) on the $\lambda=$ 850~$\mu$m continuum emission (grayscale).  
All the contours are drawn with an RMS image noise levels of 10$\sigma\times 2^n$ ($n = 0, 1, 2, 3,$ and 4) where $\sigma =$ 3.83 and 716 mJy beam$^{-1}$ for the $\lambda=$ 6\,cm and $\lambda=$ 37.1 $\mu$m images, respectively.
Overlays of (a) the $\lambda = 6$ cm continuum (contours) and (b) $\lambda=$  37.1 $\mu$m (contours) on the $\lambda=$ 850~$\mu$m continuum emission (grayscale).  
All the contours are drawn with an RMS image noise levels of $10\sigma \times 2^n$ ($n = 0, 1, 2, 3,$ and 4), where $\sigma = 3.83$ and 716 mJy beam$^{-1}$ for the $\lambda=$ 6 cm and $\lambda=$ 37.1 $\mu$m images, respectively.
The noise level of the $\lambda=$ 850 $\mu$m image is 158 mJy arcsec$^{-2}$, and the horizontal wedge shows the $\lambda=$ 850 $\mu$m intensity.
The thick-yellow contour is the $20\sigma$-level one of the $\lambda=$ 37.1 $\mu$m emission, by which we defined the area tracing the CND with contamination from the Minispiral (see \S\ref{sss:ContImages}).
The white ellipses with labels at the bottom-right corners show the beam sizes of the images (see Table \ref{tbl:Obs} for their sizes).
The horizontal bar at the top-right corner of each panel represents the projected linear size scale of 1 pc at the distance ($d$) of 8.178 kpc \citep{GRAVITY2019}.
The central cross shows the position of the Sgr A* (R.A.$ = 17^{\rm h}45^{\rm m}40^{\rm s}.0409$, Decl.$ = -29^\circ 00' 28\farcs 118$ in J2000) taken from \citet{Reid2004}.

\label{fig:ContImages}
}
\end{figure}

In Figure \ref{fig:ContImages}, we present the $\lambda = 6$~cm and 37.1\,$\mu$m continuum emission maps, using the $\lambda=$ 850\,$\mu$m image as a reference.
The overall morphology of the images across the three bands is nearly identical, with the emission being extended along Galactic Longitude and being brightest at the center where Sgr\,A is located \citep{Reid2004}.
All three wavelength datasets show prominent peaks toward the position of Sgr\,A*.
Given the beam sizes (Table \ref{tbl:Obs}) and their absolute-position accuracies, we conclude that the positions are in agreement within the errors.

The second-brightest emission in all the three bands is seen to the south-southwest of Sgr\,A*, corresponding to the Minispiral \citep{KYLo83, vol00, nit20}. 
In addition, we recognize the third-brightest $\lambda=$ 850 $\mu$m emission to the northeast whose position, although extended, is almost symmetric with respect to the position of Sgr\,A*.
We consider that the $\lambda=$ 850 $\mu$m emission morphology is reasonably described as an ellipse as a first-order approximation.
Because the 37.1 $\mu$m image has a higher angular resolution than that at $\lambda=$ 850 microns, we defined the ellipse by the $20 \sigma$-level contour of the 37.1 $\mu$m image (see Figure \ref{fig:ContImages}b).
We measure the area, $A$, enclosed by the contour to be 7750 arcsec$^2$, corresponding to $A = 12.2$ pc$^{-2}$.
This $A$ value yields an effective radius $R_{\rm eff}$ of 2.0\,pc, which is defined by $R_{\rm eff}=\sqrt{A/\pi}$.  
The center of the ellipse and axis sizes reasonably match those of the circumnuclear disk (CND) measured in previous studies \citep{gus87, chr05, req12, Hsieh2021} around Sgr\,A*, although the CND is not an isolated structure.
In conclusion, we argue that the $\lambda=$ 850 $\mu$m observations principally trace the CND with contamination from the Minispiral to the southwest.

\subsubsection{Spectral-indices Comparisons}
\label{sss:AlphaComp}
To assess the extent to which our new data sets trace the CND, a clue can be obtained from previous continuum spectrum studies \citep{Pierce-Price2000, gar11}.
\citet{gar11} derived a spectral index, $\alpha$, map (see their Figure 9) by adopting the observed intensities, $S_\nu$, at frequency, $\nu$, between 350 GHz ($\lambda=$ 850 $\mu$m) and 670 GHz ($\lambda=$ 450 $\mu$m) measured with SCUBA, and described it as $S_\nu\propto \nu^\alpha$.
According to their analysis, the CND is a low-$\alpha$ region with $-0.6 \lesssim \alpha \lesssim 1.0$ compared to the associated 20 and 50 km s$^{-1}$ clouds.\par

We compared their Figure 9 with our images (Figure \ref{fig:ContImages}) visually using graphic software and confirmed that the elliptical region (defined in \S\ref{sss:MorComp}) agrees with the low-$\alpha$ region.
Assuming that $\alpha =+3$ represents thermal emission from dust, and $\alpha = -1$ represents synchrotron emission, \citet{gar11} explained that $\alpha = +1$ is attributed to equal contributions from synchrotron and dust emission, while $\alpha = +2$ consists of 30\% synchrotron and 70\% dust emission.
\citet{gar11} also pointed out that the $\alpha$ values decrease radially toward Sgr A*, forming a halo structure of $\alpha$ over the inner CND. This can be explained by a combination of $\alpha \simeq -0.75$ synchrotron (20\% -- 45\% contribution) and $\alpha \simeq 3.0$ dust emission (80\% -- 55\%).\par

Considering the morphological comparisons (\S\ref{sss:MorComp}), the continuity of the $\alpha$ structure, and the uncertainties in the $\alpha$ analysis \citep{gar11}, we argue that the newly imaged $\lambda=$ 850 $\mu$m image primarily represents the dust emission with contamination from synchrotron emission. Specifically, the contamination ratio would range up to 45\% in the inner region and 30\% over the outer region of the CND.

\subsubsection{$\lambda=$ 850 $\mathrm{\mu}$m polarization-fraction map}
\label{ss:P850}

In Figure \ref{fig:PinCNDaround}a, we present an overlay of the polarization map at $\lambda=$ 850 $\mu$m, $P_{850}$, on the 37.1 $\mu$m emission to examine the relationship between the $P_{850}$ values and the CND.
The grayscale 37.1 $\mu$m image captures the well-defined CND and spiral structure \citep{lau13} inside the yellow contour, which we used to define the CND (\S\ref{sss:MorComp}).
Visual inspection reveals that the polarization angles in Figure \ref{fig:PinCNDaround}a display a coherent structure over the CND, which will be described in detail later in \S\ref{ss:MFresultsAnalysis}.\par

Regarding the $P_{850}$ values, we identify three remarkable features in Figure \ref{fig:PinCNDaround}a. 
First, the $P_{850}$ values increase toward the position of Sgr A*, ranging from $2.3\% \leq P_{850} \leq 4.2\%$ with a mean of $\langle P_{850} \rangle = 3.3\%$ for the four segments.
These high $P_{850}$ values are consistent with the discussion that the $\lambda=$ 850-micron emission toward Sgr A* is attributed to synchrotron emission (\S\ref{sss:AlphaComp}).
Second, except for the four segments, the polarization fractions are on the order of 0.1-1\%, which is commonly measured in the Galactic star-forming clouds (e.g, \cite{Tram22} and references therein).
We also note that the $P_{850}$ values in the northern region are slightly higher than those in the south, a feature we refer to as the N-S $P_{850}$ asymmetry.

To assess the second point above, we created Figure \ref{fig:PinCNDaround}b, which shows the radial dependence of the $P_{850}$ values (Table \ref{tbl:POL2Results}).
Note that the radius is the projected one to the plane of the sky, $r_{\rm pos}$.
We observe that the $P_{850}$ values, except for the central four points, increase outward, peak around $r_{\rm pos} \simeq 2.0$ pc, and then decrease.
Based on Figure \ref{fig:PinCNDaround}a, we emphasize that the increasing trend persists regardless of the N-S $P_{850}$ asymmetry; although the asymmetry may increase the scatter in Figure \ref{fig:PinCNDaround}b, it does not affect our findings.
The monotonic increase between $r_{\rm pos} = 0.5$\,pc and 2.0\,pc is characterized by a slope of ${\rm d}P_{850}/{\rm d}r_{\rm pos} = +0.56\pm0.26$ percent pc$^{-1}$ with Spearman's correlation coefficient (c.c.) of 0.45. 
We also find that the outer-decreasing part may have a steeper slope of ${\rm d}P_{850}/{\rm d}r_{\rm pos} = -1.0\pm 0.42$ percent pc$^{-1}$ (c.c.$=-0.47$), although the statistics are limited. 
Recall that we defined $R_{\rm eff}$ of 2.0 pc in \S\ref{sss:MorComp}, and $r_{\rm pos} \simeq 2.0$ pc is the break point of the radial profile of $P_{850}(r)$.
The correspondence in $R_{\rm eff}$, derived from $\lambda =37.1 \mu$ m data, and 
{the local peak of polarization degree $P_{850}$} suggest that 
$r_{\rm pos} \simeq 2.0$ pc could represent some characteristic scale of the CND. 

Given that the explanation for the radial profile of the $P_{850}$ values (Figure \ref{fig:PinCNDaround}b) is not straightforward, we only note that it could result from a combination of two factors. 
One is geometrical depolarization within the JCMT's beam, and the other is a misalignment and disruption of the grains, as discussed by \citet{aks2023}.
Given that the main focus of this study is the $\vec{B}$ field structure, and in light of the detailed discussion by \citet{aks2023}, we restrict our discussion to the $P$-value structure, for which multi-wavelength homogeneous polarization data are essential.

\subsection{Molecular gas in the CND traced by CS $J~=~2-1$ transition}
\label{ss:CS21}

Our objective is to investigate the \(\vec{B}\)-field structure through the newly acquired $\lambda=$ 850 \(\mu\)m polarization data, which necessitates analyzing the velocity structure of the molecular gas associated with the Circumnuclear Disk (CND), as traced by the CS \(J=2\)--\(1\) emission. 
To achieve this, we assess the gas properties; our analysis includes morphological comparisons with the continuum maps (\S\ref{sss:CS21andContMaps}) and the delineation of the CS gas linked to the CND through its centroid-velocity structure, evaluated against a toy model with spectra (\S\ref{sss:VF} and \S\ref{sss:sp}).

\subsubsection{Morphological Comparisons with the continuum maps}
\label{sss:CS21andContMaps}

Figure \ref{fig:CS21on850and37} compares the spatial extent of the CS $J=2-1$ emission with the $\lambda=$ 850 and 37.1 $\mu$m continuum maps. 
Notice that the CS map is free from the spatial-filtering effect of the interferometric observations because we utilized the ALMA's TP data.
To comprehend the overall distribution of the CS emission, we made the integrated-intensity map by calculating the zeroth order of the momentum over an LSR-velocity range of $-150 \leq V_{\rm LSR}/[{\rm km~s^{-1}}] \leq 150$, setting a detection threshold to be the 2$\sigma$ level.
Here, we present the spectra of the CS emission later in Figure \ref{fig:CS21spGuideMap}

We discern two prominent characteristics in the CS map: 
First, the molecular gas satisfying the excitation conditions for this transition extends further than the CND as outlined by the 37.1 \(\mu\)m continuum. 
Second, in contrast to the continuum maps, the emission line lacks a clearly defined elliptical shape.
Therefore, a robust comparison was conducted with the $\alpha$ map using graphical software (refer to \S\ref{sss:AlphaComp}). Visual inspection suggests that the region with low $\alpha$ values, centered around Sgr A* where synchrotron emission dominates over thermal dust emission, corresponds to the non-detection area. 
Such a morphology implies that the SMBH's 
activity, 
{along with the strong UV radiation from the rich nuclear stellar cluster surrounding the SMBH}, may have depleted molecular gas in the region.

\subsubsection{Velocity Field and Dimensions of the CND}
\label{sss:VF}

The upper-middle panel of Figure \ref{fig:CS21spGuideMap} presents the centroid-velocity map for the CS $J=2$--1 emission, derived as an intensity-weighted mean velocity map. 
A \(2\sigma\) detection threshold was used for each velocity channel to compute the zeroth and first order moment maps. 
Visual confirmation revealed that all spatio-velocity structures are consistent with previously derived centroid-velocity maps from interferometric HCN $J=1$--0 line observations \citep{gus87}, as well as HCN and HCO$^{+}$ lines \citep{chr05}. 
These authors have interpreted these spatio-velocity structures as indicative of gas rotation within the CND.
A major caveat of the CS $J=2$--1 emission data is the challenge in distinctly identifying emission originating from the CND. 
One approach to isolate the CND-originated gas is by discriminating the detected emission based on their LSR velocities.

Given the robust consistency with previous studies and the objectives of our work, we adopted the rotating-disk hypothesis to identify the velocity components attributable to the CND.
According to \citet{gus87, chr05, tsu18}, the inclination-corrected LSR velocity of the CND at each sky position for a radius of the disk can be expressed as
\begin{equation} 
V_{\mathrm{LSR}}(\Phi) = V_{\mathrm{rot}} \sin (i_{\mathrm{cnd}}) \cos(\Phi - \Phi_0) + V_{\mathrm{random}}, 
\label{eqn:Vrot}
\end{equation}
where $V_{\mathrm{rot}}$ represents the rotational velocity, $i_{\mathrm{cnd}}$ is the inclination angle relative to the line of sight (LOS), $\Phi$ is the azimuthal angle measured in a counterclockwise direction (see the bottom-right corner of Figure \ref{fig:CS21spGuideMap}), $\Phi_0$ represents the position angle of the major axis of the CND in the Galactic coordinates, and $V_{\mathrm{random}}$ denotes the ``random velocity" component, whose presence was suggested by \citet{chr05} and later confirmed by \citet{tsu18}.
Here, we adopt $V_{\mathrm{rot}} = 105\,\mathrm{km \: s^{-1}}$, $i_{\mathrm{cnd}} = 67^{\circ}$, and $\Phi_0 = 91^{\circ}$ based on the preceding argument.
Subsequently, we calculated the line-of-sight velocity using Eq.(\ref{eqn:Vrot}), varying the radius to account for projection effects. 
Because \citet{chr05} and \citet{tsu18} suggested that the ``random velocities" vary between 10 and 30 $\mathrm{km\:s^{-1}}$ in the LSR velocity frame across the CND, we included a conservative estimate of the maximum value of $\pm 30$ $\mathrm{km\:s^{-1}}$ to calculate the $V_{\mathrm{LSR}}$ at each sky position.

To facilitate the analysis of the velocity structure and $\vec{B}$-field structure (to be detailed in \S\ref{sss:MFresults} with Figure~\ref{fig:Bmaps}), as well as to examine the known structures, we divided the ring into five subregions, labeled A to E, as depicted in the lower-middle panel. 
Subregion B aligns closely with the Southwest lobe, while subregions E and D correspond to the Northern Arm and Northeast Lobe, respectively. 
Subregions B, C, and D overlap with the Western Arc of the Minispiral, as traced by the $\lambda = 6$~cm and $\lambda = 37.1\:\mu$m continuum emission maps (Figure \ref{fig:ContImages}). 
The first three rows of Table \ref{tbl:Regions} summarize the resultant ranges of the desired $V_{\mathrm{LSR}}$, employed to isolate gas originating from the CND, and the $\Phi$ ranges within which the subregions were arbitrarily defined.

A robust comparison between the observed and the model centroid-velocity maps indicates that the simple model described by Eq.(\ref{eqn:Vrot}) is reasonably effective. 
It captures the presence of blueshifted gas on the southern side of the ring and redshifted gas to the north. 
However, the observed velocity structure is complex, likely influenced by the infall \citep{mon09} and turbulence, as indicated by the $V_{\rm random}$ term. 
For instance, a steep velocity gradient ($V_{\rm LSR} \simeq +50$ km s$^{-1}$ to $-50$ km s$^{-1}$ over approximately 1\,pc) was observed immediately inside subregion B, where it overlaps with the Western Arc.

Lastly, we present the estimates of the dimensions of the CND.
For the centroid-velocity model map shown in Figure \ref{fig:CS21spGuideMap}, we modeled the inner ellipse with major and minor axes of $r^{\rm inner}_{\rm maj} = 1.5$\,pc and $r^{\rm inner}_{\rm min} = 0.5$\,pc, and the outer ellipse with $r^{\rm outer}_{\rm maj} = 2.5$\,pc and $r^{\rm outer}_{\rm min} = 1.5$\,pc, respectively.
The inner radii and the disk-inclination angle set a radius of the inner cavity $r_{\rm cav}^{\rm 3D} \simeq 2.2$\,pc through $\sqrt{r^{\rm inner}_{\rm maj} \times r^{\rm inner}_{\rm min}}/\cos(67^\circ)$. 
This estimate, based on the 2\farcs6 resolution CS $J=2-1$ line, is larger than the 1.4 pc derived from the 4\farcs6 resolution 37.1 $\mu$m emission imaging by \citet{lau13}.
We argue that this discrepancy could be attributed to the differences in tracers and angular resolutions.
The outer-axis sizes give an effective-outer radius of $(2.5\times 1.5)^{1/2}$\,pc $\simeq$1.9\,pc, yielding a 3-dimensional CND radius of $R_{\rm eff}^{\rm 3D} \simeq 5.0$\,pc. 
Note that the inferred 1.9\,pc aligns with the projected-border radius of 2.0\,pc between the magenta-and-green symbols in Figure \ref{fig:PinCNDaround}b where the radial-$P_{850}$ profile changes.

\subsubsection{The Spectra in Individual Subregions}
\label{sss:sp}
The associated spectra in Figure \ref{fig:CS21spGuideMap} were obtained by averaging the pixel values over each subregion regardless of the presence or absence of signals.
All the CS spectra exhibit complicated spectral shapes. 
In the velocity range encompassed by the vertical dotted line, the spectrum obtained in subregion A shows a negative dip around $V_{\rm LSR} = 22.0$ km s$^{-1}$, and the spectrum from subregion B displays a local minimum at the same velocity, suggesting the presence of a common foreground cold gas at that velocity.
Indeed, previous studies have shown that there exist broad absorption features at $V_{\rm LSR} = -55$ km s$^{-1}$ (due to the 3 kpc arm), $-30$ km s$^{-1}$ (unknown), and 0 km s$^{-1}$ (local) \citep{gus87, Herrnstein2002, {Sor13}} caused by extended foreground cloud complexes.\par

These spectra present non-negligible caveats for deriving meaningful physical quantities due to the presence of multiple components and absorption features. This complexity suggests the coexistence of multiple gas components along the line of sight, combined with self-absorption and/or absorption by intervening gas.
Such complexities are commonplace in molecular lines observed towards the Galactic Center.
Although the LSR-velocity ranges defined by the ring model are evident, these caveats significantly impede further analysis.
Consequently, we constrain our study to utilize the CS lines to define the subregions.

\subsection{$\vec{B}$-field structure}
\label{ss:MFresultsAnalysis}

\subsubsection{$\vec{B}$-field structure inferred from the $\lambda=$ 850 $\mu$m Observations}
\label{sss:MFresults}

Figure~\ref{fig:Bmaps} presents the \(\vec{B}\)-field structure map, derived by rotating the polarization segments in Figure~\ref{fig:PinCNDaround} by \(90^\circ\). 
The 12\farcs6 resolution map distinctly reveals a 1-pc scale \(\vec{B}\)-field structure within the CND, exhibiting coherent \(\vec{B}\)-field lines with subtle directional variations. 
Excluding four prominent segments directed toward Sgr~A*, the overall geometry predominantly resembles a gently winding spiral centered on the radio source, displaying point symmetry around Sgr~A*. 
On the western side of the CND, the $\vec{B}$-field directions align with the CND ring, while on the eastern and northern sides, they diverge from the CND ring's path. 
Notably, the observed the $\vec{B}$-field directions consistently parallel the minispiral's orientation.
One should recall that the $\lambda=$ 850 $\mu$m observations are more likely to reflect the warm molecular component of the CND gas rather than the Minispiral, which consists of hot ionized gas, as indicated by the low $P_{850}$ values in the radial plot (Figure \ref{fig:PinCNDaround}).
The geometry of the \(\vec{B}\) field will be further analyzed in \S\ref{sss:WKmodel}.

\subsubsection{Analysis of Magnetic-field Structure}
\label{sss:WKmodel}

In this subsection, we delineate our comparative analysis between the observed $\vec{B}$ field (Figure \ref{fig:Bmaps}) and the 
\(\vec{B}\)  field models posited by \citet{war90}, henceforth referred to as the WK model. 
The WK model, encompassing five variations denoted as gc1, gc2, gc3, gc4, and gc5, proposes that 
{$\vec{B}$-field lines}, initially perpendicular to the Galactic plane, are swept into a rotating radial gas inflow, which results in the $\vec{B}$ field  lines in the Galactic plane becoming aligned, eventually parallel to the plane. 
The fundamental assumption of this model is that the excess angular momentum of the accreting material is transported by magnetic stresses, thereby obviating the need to invoke viscosity.
The WK model has been acclaimed for accurately depicting the $\vec{B}$ field  structure at the Galactic center, corroborated by observations at 40$^{\prime\prime}$ resolution in 100 $\mu$m polarimetry from the Kuiper Airborne Observatory \citep{hil90}, and at 20$^{\prime\prime}$ resolution in $\lambda=$ 850 $\mu$m observations with SCUPOL \citep{Hsieh2018}.

{The five models are described in cylindrical coordinates, characterizing the radial, azimuthal, and vertical components of the $\vec{B}$ field as \(B_r\), \(B_{\phi}\), and \(B_z\), respectively\footnote{When we apply these models, we set the angle between the CND's major axis and the galactic latitude axis as 2.5 degrees, which can be derived based on the angle between the R.A. axis and the CND's major axis of 25 degrees \citep{Sato2008} and the angle between the equatorial coordinate system and the galactic coordinate system of 62.5 degree.}.} 
Since \citet{war90} did not solve the full set of dynamical equations, they introduced four dimensionless free parameters: \(\alpha_{\rm WK}\), \(\beta_{\rm WK}\), \(\delta_{\rm WK}\), and \(\varepsilon_{\rm WK}\). 
According to \citet{kon89, war90}, the parameter \(\alpha_{\rm WK}\) represents the ratio of the $\vec{B}$-field strengths between the azimuthal and vertical components, \(\frac{|B_{\phi}|}{|B_z|}\). Likewise, \(\beta_{\rm WK}\) denotes the ratio between the radial and vertical components, \(\frac{|B_r|}{|B_z|}\), leading to the definition of another useful parameter \(\eta\equiv\frac{\beta_{\rm WK}}{\alpha_{\rm WK}}\), or \(\eta = \frac{B_r}{B_{\phi}}\). 
The parameter \(\delta_{\rm WK}\) is defined as the ratio of the disk's half-thickness to its radius, with the assumption that \(\delta_{\rm WK} \ll 1\). 
The parameter \(\varepsilon_{\rm WK}\) represents the ratio of the inflow velocity to the azimuthal velocity, \(\left|\frac{v_r}{v_{\phi}}\right|\).
Although these four parameters are presented as free parameters, fiducial values are provided in the \citet{war90} paper (refer to their Table 1).

The parameters \(\alpha_{\rm WK}\), \(\beta_{\rm WK}\), \(\delta_{\rm WK}\), and \(\varepsilon_{\rm WK}\) should be updated with measurements taken post-publication of \citet{war90} and assessed, as detailed in \S\ref{s:D}.
In practice, the selection of values for \(\alpha_{\rm WK}\), \(\beta_{\rm WK}\), \(\delta_{\rm WK}\), and \(\varepsilon_{\rm WK}\) does not influence the calculations of the Stokes $q$ and $u$ parameters. 
Using the angles \(\theta\) and \(\omega\), which \citet{war90} employed to explain the $\vec{B}$-field geometry (see their Figure 4b), their Eqs.(3.20) and (3.21) at a disk position described by the azimuthal angle \(\phi\) can be respectively rewritten as
\begin{equation}
u(\theta, \omega) = N_d \,\mathrm{cos}\,i_{\mathrm{disk}} \,\mathrm{sin}^2 \omega \,\mathrm{sin}\left\{2(\theta+\phi)\right\},
\end{equation}
and
\begin{eqnarray}
q(\theta, \omega) &=& N_d \,\mathrm{sin}^2 \omega \left[\mathrm{cos}^2\left\{2(\theta+\phi)\right\} \, (\mathrm{cos}^2 i_{\mathrm{disk}} + 1) - 1\right] \nonumber \\
 &+& \mathrm{cos}^2 \omega \,\mathrm{sin}^2 i_{\mathrm{disk}},
\end{eqnarray}
which are independent of the \(\alpha_{\rm WK}\), \(\beta_{\rm WK}\), \(\delta_{\rm WK}\), and \(\varepsilon_{\rm WK}\) parameters, even when considering them as arguments in implicit functions. 
Here, \(N_d\) denotes the column density of dust, and \(i_{\rm disk}\) is the inclination angle of the disk with respect to the line of sight. 
The angles \(\omega\) and \(\theta+\phi\) [the sum of which is denoted as \(\xi\) in \citet{war90}] define the \(\vec{B}\)-field geometry. 
\(\omega\) is, as illustrated in Figure 4b of WK model paper \citep{war90}, the angle between the \(\vec{B}\)-field line and the \(z\)-axis, and \(\theta\) is the angle between the (cylindrical) radial vector and the projected angle of the upper layer \(\vec{B}\) field vector onto the mid-plane of the disk. 

We adopted the disk inclination angle \(i_{\mathrm{disk}} = 67^{\circ}\) from \citet{lau13} who reported the value with an error of $5^{\circ}$, consistent of $70^{\circ}$ by \citet{gus87}, along with the \(\omega\) and \(\theta\) values from Table 1 of \citet{war90} for the five models.
This approach consequently renders the Stokes \(u\) and \(q\) as functions of a single parameter, \(\phi\). 
Therefore, the desired position angle of the \(\vec{B}\)-field line at any given radial outward line can be calculated by \(PA_{\rm model} = \frac{1}{2}\arctan\left(\frac{u}{q}\right) + \frac{\pi}{2}\).
In practice, we measured the $\phi$ value at each position and calculated the Stokes $q$ and $u$ parameters to determine the $PA_{\rm model}$ for each position. 
The $\vec{B}$ field structures were computed according to the five models detailed in Table 1 of \citet{war90}. 
The $\chi^2$ and the ``Residual" columns in Table \ref{tbl:model_fits} summarize the outcomes of the model fitting, and Figure \ref{fig:GC123models} illustrates comparisons of the $\vec{B}$-field structures between the observations and the gc1, gc2, and gc3 models, which showed better fits compared to gc4 and gc5.
The goodness of each fit was evaluated using residuals, defined as $\Delta PA \equiv PA_{\mathrm{model}} - PA_{\mathrm{obs}}$.
The quality of the fits was quantified with a Chi-square fit, given by $\chi^2=\Sigma(\Delta PA)^2/n$, where $n$ represents the number of segments.
Since our analysis is focused on the CND, we limited our consideration to $\vec{B}$-field segments within the 20$\sigma$-level contour of the 37.1 $\mu$m emission. 
We excluded the four magenta-coded segments identified as being affected by Sgr A* (see Figure \ref{fig:PinCNDaround}), resulting in a total of $n=51$ segments for analysis.

Our results are essentially consistent with those reported by \citet{war90}, who investigated the 100 $\mu$m polarization data from the KAO \citep{hil90}. 
We found that the gc1, gc2, and gc3 solution series explain the results more effectively than the remaining two models.
Although the $\chi^2$ values of these three models are comparable, the gc1 model is preferred due to its minimal median absolute deviation (MAD$_{\rm stat}$) in residuals.
Here, we selected the MAD$_{\rm stat}$ as our metric of choice, rather than the median, due to the limited sample size of $n=51$. 
MAD$_{\rm stat}$ is particularly suited for small datasets as it provides a more robust measure against the outliers that can skew the results. 
In contrast, the mean is not an appropriate metric in this context because of its sensitivity to extreme values.

The distinctions between these models are evident in Figure 2 of \citet{war90} (see also Figure 3 in \cite{Hildebrand1993}), which illustrates the relative importance of the $B_r$, $|-B_\phi|$, and $B_z$ components along the vertical direction, $z/r$. 
The thickness of the CND is estimated to be approximately 0.5 pc at $r = 2$ pc \citep{gus87} and 0.34\,pc at $r =$ 1.4\,pc \citep{lau13}. 
We, therefore, arbitrarily adopted a geometrical mean of 0.4\,pc as the half-thickness of the disk, $z_{\rm half}$. 
Together with the $R_{\rm eff}^{\rm 3D}$ value in \S\ref{sss:VF}, we obtained a ratio of 
\begin{equation}
\frac{z}{r} = \frac{z_{\rm half}}{R_{\rm eff}^{\rm 3D}} \simeq  0.08
\label{eq:delta_WK}
\end{equation}
with an uncertainty of approximately 30\%. 
This ratio corresponds to the \(\delta_{\rm WK}\) parameter as given in Eq.(2.2c) of \citet{war90}, thereby satisfying the thin-disk approximation.

Consequently, we infer that Figure 2 of \citet{war90} depicts the relative importance of the $B_r$, $|-B_{\phi}|$, and $B_z$ components within the CND.
In the gc1-3 models, the contributions from the radial and azimuthal components vary along the vertical direction: $|-B_{\phi}| > B_r$ at low altitudes near the disk equatorial plane where $z/r\lesssim 0.1$, while the opposite is true in the high altitude regions. 
This leads to an overall spiral pattern, as observed in Figure \ref{fig:GC123models}. 
In contrast, the gc4 model, which is almost toroidal, and the gc5 model are characterized by an azimuthal $\vec{B}$ field within the disk. 
When assessing the effectiveness of the models based on the $\chi^2$ and MAD$_{\rm stat}$ values, the $|-B_\phi|$-dominated models appear to be less successful.

Thus far, we have assessed the overall goodness of the five models by analyzing their $\chi^2$ and MAD$_{\rm stat}$ values, as detailed in Table \ref{tbl:model_fits}. 
Focusing on the maps in Figure \ref{fig:GC123models}, our comparison is specifically concentrated within the CND, demarcated by the yellow contour. 
Remarkably, the three models (gc1, gc2, and gc3) successfully replicate the $\vec{B}$-field directions in the Western Arc's brightest region of the Minispiral. 
The gc1 model, and to a certain degree gc2, reasonably depict the observed $\vec{B}$-field on the left (north-north-eastern) side of Sgr A*. 
However, the gc3 model exhibits more significant deviations from the observed data. 
Along the ring's minor axis, only gc1 accurately reflects the observations, particularly on the map's lower (south-south-western) side.
It is noteworthy that all models, including gc4 and gc5 
(See {Appendix} Figure \ref{fig:gc4gc5}), are unable to replicate the observed angles in the Subregion A  identified in the CS line velocity field analysis (see Figure \ref{fig:CS21spGuideMap}).

\section{Discussion}
\label{s:D}
Our analysis of the newly obtained polarization data has revealed the presence of a well-ordered $\vec{B}$ field across the CND. 
With a linear resolution of 0.47\,pc, the analysis indicates that the observed \(\vec{B}\)-field structure is consistently reproduced by the accretion disk model, which posits that the disk is threaded by open $\vec{B}$-field lines, as discussed in \S\ref{s:ResAna}. 
Subsequently we estimate the $\vec{B}$-field strengths, \( |B| \), in \S\ref{ss:Bvalue}, and subsequently discuss their implications in \S\ref{ss:Role_of_B}.

\subsection{Estimate of \( |\vec{B}| \) values based on the \(\lambda = 850 \mu\)m polarization observations}
\label{ss:Bvalue}
In this subsection, we aim to establish observational constraints on the $\vec{B}$-field strength, \( |B_0| \), at the center of the inner-ionized cavity (\S\ref{sss:B0}). 
The determination of \( |B_0| \) enables us to estimate the field strength within the surrounding CND (\S\ref{sss:BrBphi}).

\subsubsection{Estimating the Magnetic-field strength at the Midplane Center of the Inner Ionized Cavity, $|B_0|$}
\label{sss:B0}
In our analysis of the \(\vec{B}\)-field structure, we utilized the \citet{war90} model, as elaborated in \S\ref{sss:WKmodel}. 
The sole free parameter in our model application was the azimuthal angle \(\phi\) in the disk, with other parameters set to fiducial values from Table 1 of \citet{war90}.
Therefore, we revisited the \(\alpha_{\rm WK}\) and \(\beta_{\rm WK}\) parameters, which represent the three-dimensional magnetic-field strength over the inner ionized cavity of the disk, indicated by \(|B_0^{\rm 3D}|\). 
{Note that the subscript 0 stands for the midplane values.} 
To determine these parameters, updated measurements for each quantity were considered, except for \(|B_0^{\rm 3D}|\). 
We aimed to ensure that the resultant \(\alpha_{\rm WK}\) and \(\beta_{\rm WK}\) values would be within the ranges of \(15\lesssim \alpha_{\rm WK}\lesssim 40\) and \(17\lesssim \beta_{\rm WK}\lesssim 27\), as specified for the gc1, gc2, and gc3 models in Table 1 of \citet{war90}.

In practice, we can consider Eqs~(2.11a) and (2.11b) of \citet{war90} to be functions of $B_0$ for $\alpha$ and $\beta$, respectively, once other parameters are fixed. 
Because the dependency on $B_0$ is common in both $\alpha(B_0)$ and $\beta(B_0)$, it is impossible to have a $B_0$ value that satisfies both functions simultaneously.
Therefore, we numerically searched for the ``most plausible $B_0$'' value, considering the average of $\alpha(B_0)$ and $\beta(B_0)$, so that they approximate the $\alpha$ and $\beta$ values given in Table 1 of \citet{war90} as closely as possible. 
The resultant $\alpha$, $\beta$, and $B_0$ values are summarized in Table \ref{tbl:model_fits}. 
As can be seen in the $|B_0^{\rm 3D}|$ column, the order of the model corresponds to the order of the magnetic-field strengths. 
Considering the $\chi^2$ values, we also presented the weighted-mean strength of $0.24^{+0.05}_{-0.04}$ mG calculated for the gc1, 2, and 3 models, which is used in subsequent analysis and referred to as $|B_0^{\rm 3D}|$. 
Here, the asymmetry in the errors is attributable to the use of a weighted average.

Using the best-estimated values, we can rewrite Eq.~(2.11a) of \citet{war90} as,
\begin{eqnarray} 
\alpha_{\rm WK} \simeq 32 \left(\frac{n_{\rm H}}{2.5\times 10^3\, {\rm cm}^{-3}}\right) \left(\frac{v_{\phi0}}{105\, {\rm km\,s}^{-1}}\right)^2 \nonumber \\
\left(\frac{\epsilon_{\rm WK}}{0.22}\right) \left(\frac{|B_0^{\rm 3D}|}{0.24^{+0.05}_{-0.04} \, {\rm mG}}\right)^{-2}.
\label{eqn:alpha_WK}
\end{eqnarray}
where $n_{\rm H}$ is the mean atomic hydrogen density over the inner cavity, $v_{\phi0}$ the midplane rotational velocity of the CND from our analysis, and $\epsilon_{\rm WK}$ the ratio of inflow to azimuthal velocity.
The difficulty in selecting an $n_{\rm H}$ value stems mainly from observational challenges. 
\citet{war90} used the lower limit of $n_{\rm H} \gtrsim 10^4$ cm$^{-3}$ from [OI] 63 $\mu$m spectra \citep{gen85}, while multi-wavelength continuum imaging at JCMT suggestd $n_{\rm H} \simeq 2.5\times 10^3$ cm$^{-3}$ \citep{zyl95}, although theoretical studies suggest lower densities [e.g., \citet{bla16}]. 
Considering the ambiguity of the observational estimate, we selected the density estimated from the continuum spectrum analysis.
Notice that the estimate by \citet{Zhao2010} toward the ``minicavity'' does not spatially match with the inner cavity.
Similarly, we estimate a \(|B_{0}^{\rm 3D}|\) value that aligns with the \(\beta_{\rm WK}\) range by revising Eq.(2.11b) from \citet{war90} as follows:
\begin{eqnarray} 
\beta_{\rm WK} \simeq 17 \left(\frac{n_{\rm H}}{2.5\times 10^3\, {\rm cm^{-3}}}\right)^{3/2} \left(\frac{v_{\phi0}}{105\, {\rm km~s^{-1}}}\right)^2 \nonumber \\
\left(\frac{\epsilon_{\rm WK}}{0.22}\right) \left(\frac{r_{\rm cav}^{\rm 3D}}{2.2~{\rm pc}}\right) \left(\frac{|B_0^{\rm 3D}|}{0.24^{+0.05}_{-0.04} ~{\rm mG}}\right)^{-2}
\label{eqn:beta_WK}
\end{eqnarray}
where \( r_{\rm cav}^{\rm 3D} \) represents the midplane radius of the ionized cavity (\S\ref{sss:VF}).

\subsubsection{Estimating the $\vec{B}$-Field Strength in the CND and Its Implications}
\label{sss:BrBphi}

In the previous subsection, we estimated the $\vec{B}$-field strength at the midplane center of the ionized cavity, \(|B_0^{\rm 3D}|\), which corresponds to $B_0$ in \citet{war90}. 
These authors showed that the azimuthal and radial components of the field strengths can be approximated as $B_\phi /B_0\simeq -\alpha\frac{z}{r}$ and $B_r /B_0\simeq \beta\frac{z}{r}$ for $z\ll r$, respectively [see their Eqs.(2.5) and (2.8), along with their Figure 2 (see also Figure 3 in \cite{Hildebrand1993})]. 
Given our estimates of $z/r$, which corresponds to the \(\delta\) parameter in \citet{war90}, and the values of \(\alpha_{\rm WK}\) and \(\beta_{\rm WK}\) [see Eqs.(\ref{eq:delta_WK}), (\ref{eqn:alpha_WK}), and (\ref{eqn:beta_WK})], we can estimate the $\vec{B}$-field strength in the CND by applying these values to the equations as follows:
\begin{equation} 
B_\phi \simeq -0.7\pm 0.2 \,{\rm mG}
\left(\frac{\alpha_{\rm WK}}{34^{+10}_{-13}}\right)
\left(\frac{z/r}{0.08}\right)
\left(\frac{B_0^{3D}}{0.24^{+0.05}_{-0.04} ~{\rm mG}}\right)
\label{eqn:B_phi}
\end{equation}
and
\begin{equation} 
B_r \simeq 0.4\pm 0.1 \,{\rm mG}
\left(\frac{\beta_{\rm WK}}{18\pm 6}\right)
\left(\frac{z/r}{0.08}\right)
\left(\frac{B_0^{3D}}{0.24^{+0.05}_{-0.04} ~{\rm mG}}\right).
\label{eqn:B_r}
\end{equation}
Note that the negative sign in \(B_\phi\) is necessary to ensure that the \( -B_r B_\phi > 0 \) condition is met in the Maxwell stress term of the magnetohydrodynamic (MHD) equations, a requirement for outward angular-momentum transfer by ordered $\vec{B}$ field. 
We will revisit this aspect later in \S\ref{ss:Role_of_B}.
We derived the \(\beta_{\rm WK}\) and \(\alpha_{\rm WK}\) values from the mean of these parameters summarized in Table 1 of \citet{war90} and from Eqs.(\ref{eqn:alpha_WK}) and (\ref{eqn:beta_WK}). 
According to Figure 2 in \citet{war90}, the vertical component, $B_z$, can be approximated as $B_z \simeq f_z B_0$ where the scale factor $f_z$ remains unity up to $z/r \simeq 0.1$ for the gc1, gc2, and gc3 models. 
We, therefore, obtain the vertical component at $z/r = 0.08$,
\begin{equation} 
B_z \simeq 0.2 \pm 0.05 \,{\rm mG}
\left(\frac{f_z}{1.0}\right)
\left(\frac{B_0^{\rm 3D}}{0.24^{+0.05}_{-0.04} \,{\rm mG}}\right).
\label{eqn:B_z}
\end{equation}
A straightforward comparison of Eqs. (\ref{eqn:B_phi}), (\ref{eqn:B_r}), and (\ref{eqn:B_z}) suggests that the CND's $\vec{B}$-field has a $B_\phi$-dominated structure, indicative of a predominantly toroidal field. 
If the condition $|B_\phi| \gtrsim B_z$ holds true, it suggests the amplification of the $\vec{B}$-field in differentially rotating flows.
Assuming that the CND originally possessed a vertical $\vec{B}$ field and the gas pressure dominates over the magnetic pressure, 
MRI \citep{Vel59, cha61, BH91} would continuously drive the $B_r$ component.
Concurrently, the differential rotation within the CND would enhance the toroidal ($B_\phi$) component, a result of the rotation stretching the field lines, leading to a $B_\phi$-dominated field. 
The dominance of the toroidal component arises from the effective elongation of $\vec{B}$-field lines by the velocity shear. 
Importantly, the observation that $|B_\phi| > B_r \simeq B_z$ provides strong support for the idea that MRI indeed facilitates the transfer of excess angular momentum outward across the CND.
We emphasize that our interpretation is consistent with the assumptions made in \citet{kon89} and \citet{war90}.

After determining the three components \( B_r \), \( B_\phi \), and \( B_z \), we arrive at an estimate for the 3D $\vec{B}$-field strength of the CND, denoted as \( B_{\rm CND}^{3D} = (B_r^2 + B_\phi^2 + B_z^2)^{1/2} \). The calculated strength is:
\begin{equation} 
B_{\rm CND}^{3D} \simeq 0.8 \pm 0.3 \,{\rm mG}.
\label{eqn:B_CND}
\end{equation}
The resultant 3D strength is comparable to the representative plane-of-sky component strength over the CND, \(B_{\rm POS}^{\rm DCF,SF} \simeq 1.0\pm 0.2 \,{\rm mG}\), as derived by \citet{gue23} (see their Table 5). 
Similarly, \citet{kil92} reported a line-of-sight $\vec{B}$-field component of approximately 2 mG in the Northern part of the CND, based on the Zeeman effect observed in the OH absorption line.

Our insights prompt a reevaluation of the $\vec{B}$ field's role within the CND, potentially applicable to similar structures in other galaxies. 
Finally, updating the estimate of the mass accretion rate toward Sgr A*, 
{based on Eq. (5.3) 
in \citet{war90}}, becomes pertinent:
\begin{eqnarray} 
\frac{{\rm d}M}{{\rm d}t} &\sim& 3\times 10^{-2} \,M_\odot ~{\rm yr}^{-1}
\left(\frac{r_{\rm cav}^{\rm 3D}}{2.2~{\rm pc}}\right)^2 
\left(\frac{v_{\phi0}}{105\, {\rm km~s^{-1}}}\right)^{-1} \nonumber \\
 & & \left(\frac{B_0^{3D}}{0.24^{+0.05}_{-0.04} ~{\rm mG}}\right)^2
\left(\frac{|B_\phi|}{0.7\pm 0.2 \,{\rm mG}}\right).
\label{eqn:dMdt}
\end{eqnarray}

\subsubsection{An upper limit of a typical $|\vec{B}|$ over the CND based on the Davis–Chandrasekhar–Fermi method}
\label{sss:DCF}

Given the estimated angular deviation from the ordered \(\vec{B}\)-field lines, as indicated by the residuals from the model in Table \ref{tbl:model_fits}, we evaluate the applicability of the Davis–Chandrasekhar–Fermi (DCF) method to the CND. 
This assessment involves estimating the representative strength of the plane-of-sky $\vec{B}$-field component, \(B^{\rm pos}_{\rm CND}\). 
However, a naive application of the DCF method can be problematic in regions where magnetic energy is not in equilibrium with the kinetic energy of turbulence. 
This is particularly true in the innermost, densest regions of clouds where self-gravity predominates over gas motions.

In the specific case of the CND, where magnetic-field lines are presumed aligned by the streamers of ionized materials accreting onto Sgr A*, applying the DCF method in its original form may not yield an accurate estimate of the plane-of-sky $\vec{B}$-field strength, \( B^{\rm pos}_{\rm CND} \) \citep{gue23}. 
\citet{gue23} proposed a modified DCF method, incorporating the ideal magnetohydrodynamic (MHD) equations and accounting for the effects of differential rotation and shear flow (SF) within the CND. 
This approach yielded an estimate of \( B^{\rm pos, sf}_{\rm CND} \), denoted as \( B^{\rm DCF, SF}_{\rm POS} \) in their notation), of 1.0\(\pm\)0.2 mG.
Although their formulation is crucial for studying the role of $\vec{B}$-field in disks like the CND, we note that the estimated values might be subject to systematic errors due to uncertainties in gas density and possibly velocity dispersion. 
For instance, using large velocity gradient (LVG) analysis on multi-transition CS line data from ALMA, \citet{tsu18} estimated the molecular hydrogen number density (\( n_{\rm H_2} \)) to be approximately \( 2.2 \times 10^5 \) cm\(^{-3}\), an order of magnitude greater than the value used by \citet{gue23}  (\( n_{\rm H_2} \simeq 1.9 \times 10^4 \) cm\(^{-3}\) converted from \( \rho \) in their Table 5).
Additionally, caution should be also exercised when referencing the LVG analysis by \citet{tsu18}, particularly regarding their assumptions about the velocity width (50 km s$^{-1}$ in FWHM), yielding a non-thermal velocity dispersion of $\sigma_{\rm nth}\approx$ 21 km s\(^{-1}\), $\approx 60\%$ of that used in \citet{gue23}, and the fraction abundance of the molecule.
Unfortunately, we stress that such a large uncertainty in the density and velocity dispersion estimates potentially outweighs the improvements offered by the modified DCF method.

Regardless of the methodological revisions or estimates, systematic errors in estimating \( B^{\rm pos}_{\rm CND} \) are inevitable. 
Consequently, we advocate for the use of the original DCF formula, derived from the equilibrium between the kinetic energy of turbulence and the magnetic energy (\( \frac{\rho}{2}v_{\rm turb}^2 = \frac{B^2}{8\pi} \)). 
The equipartition assumption yields the widely known expression \(B^{\rm pos}=\sqrt{4\pi\rho}\frac{\sigma_{\rm nth}}{\sigma_{\phi}}\), where \(\sigma_{\phi}\) denotes the dispersion in polarization angle. 
We apply this formula using two sets of estimates for density (\(\rho\)) and non-thermal velocity dispersion (\( \sigma_{\rm nth} \)) from \citet{tsu18}:
\begin{eqnarray}
B_{\rm CND}^{\rm pos} \lesssim 72\,\mathrm{mG}
\left(\frac{Q}{1.0}\right)
\left(\frac{\rho}{1.0\times 10^{-18}\,{\rm g\,cm^{-3}}}\right)^{1/2} \nonumber \\
\left(\frac{\sigma_{\rm nth}}{21\,{\rm km\,s^{-1}}}\right)
\left(\frac{\sigma_{\phi}}{0.105\,{\rm rad}}\right)^{-1}.
\label{eq:DCF_tsu18}
\end{eqnarray}
where \( Q \) is a correction factor for projecting the unknown 3D field strength onto the plane-of-sky component and accounts for beam dilution effects, with a range of \( 0.0 \leq Q \leq 1.0 \) \citep{ost01}. 
The \( \sigma_{\phi} \) value is derived from the MAD$_{\rm stat}$ of the residual for the gc1 model (Table \ref{tbl:model_fits}). 
Note that in our comparison with the results from \citet{gue23}, we did not apply the \( Q \) correction. 
By selecting \( Q = 1.0 \), Eq.(\ref{eq:DCF_tsu18}) provides an upper limit. 
This estimate should be viewed as a representative, likely mean value over the CND, consistent with the original considerations of \citet{dav51} and \citet{cha53}. 
Furthermore, when employing the \(\rho\) and \(\sigma_{\rm nth}\) values used by \citet{gue23}, we derive a tighter constraint for \(B_{\rm CND}^{\rm pos} \lesssim 35\,\mathrm{mG}\) with \(\rho = 8.9\times 10^{-20}\,{\rm g\,cm^{-3}}\) and \(\sigma_{\rm nth} = 35\,{\rm km\,s^{-1}}\).

We emphasize significant limitations in applying the DCF method to the CND. 
First, the selection of $\rho$ and $\sigma_{\rm nth}$ values is inherently arbitrary, as there is no conclusive evidence to confirm whether the molecular line observations correspond to the same gas regions captured by polarimetric-continuum imaging. 
Second, the assumption of equipartition between magnetic and kinetic energies is subject to skepticism. 
Third, even if these two points are cleared, the relatively small number of polarization segments (51 in our case) may not provide a solid basis for an accurate estimation of \(\sigma_{\rm nth}\). 
Fourth, the \citet{war90} model may not fully account for systematic gas motions, thereby making cautious of the separation of purely turbulent contributions. 
Consequently, we ascertain that the DCF method does not provide meaningful constraints on the plane-of-sky component of the $\vec{B}$-field strength in the CND.

\subsection{Importance of $\vec{B}$ Field in the CND}
\label{ss:Role_of_B}
\subsubsection{Role of $\vec{B}$ Field in sustaining the CND: An Implication from Multiple-Model Comparisons}
\label{sss:Role_of_B}

It is essential to recognize that the WK model predicts the global structure of the $\vec{B}$ field permeating a rotating gas disk, focusing on the broader configuration rather than localized structures. 
We discussed how turbulence in the magnetized gas plays a crucial role in sustaining the current $\vec{B}$-field (\S\ref{sss:BrBphi}). 
It's also essential to recognize that the observed $\vec{B}$-field direction at any given point represents a vector average along the line of sight. 
Consequently, caution is required in over-reliance on the WK model for elucidating local structures. 
Nevertheless, the prospect of employing a combination of gc1-5 models to delineate the overarching structure remains intriguing. 
Recall that the direction of the model $\vec{B}$ field at each sky position is determined by a singular parameter, $\phi$ (\S\ref{sss:WKmodel}), with all other parameters fixed for each of the gc1-5 models. 
Consequently, we chose the model that minimized the residual at each position, which led to the creation of Figure \ref{fig:GC12345models}.

Based on the fitting results presented in Table \ref{tbl:model_fits}, where models gc1-3 exhibit comparable and smaller $\chi^2$ values compared to gc4 and 5, we present two sets of multiple-model fittings in Figure \ref{fig:GC12345models}. 
The (b) illustrates the fittings from gc1-3, while the (c) panel includes gc1-5. 
It should be noted that the $\chi^2$ values were computed for the 51 segments within the yellow contour, excluding the four segments exhibiting high $P_{850}$ values (\S\ref{sss:WKmodel}).
Consequently, our discussion is focused on the segments for which $\chi^2$ values were calculated. 
Examination of these panels reveals that the large-$\alpha$ models (gc1 and 2) effectively explain the observed $\vec{B}$-field orientation. However, the strong $\vec{B}$-field model of gc3 also provides a satisfactory representation of several southern segments. 
These findings suggest that, under the assumption of the $\vec{B}$ field being frozen into the gas, the field is predominantly not significantly wound up by the rotational motion of the gas. 
The validity of model gc3 also implies that ambipolar diffusion, as hypothesized, may be locally operational, possibly in regions with low ionization degree [see Eq.(2.2b) in WK].

In light of the fact that not all observed $\vec{B}$-field orientations were reasonably reproduced by the gc1-3 models, we assessed whether the models with higher $\chi^2$ values (gc4 and 5) could locally explain the observed field directions. 
Upon comparing the right panel of Figure \ref{fig:GC12345models} with Figure \ref{fig:PinCNDaround}a, it is apparent that the red-gc5 segments correspond with areas of high polarization at the center, as indicated by the magenta color in Figure \ref{fig:PinCNDaround}a. 
Consequently, these two segments are omitted from our analysis of the CND. 
While two of the four orange segments (gc4) located to the north-north-west of Sgr A* may appear consistent with the model, considering the uncertainties inherent in both observations and modeling, we deduce that the small-$\beta$ models (gc4 and 5) are not locally applicable. 
The above result suggests a lack of significant magnetic diffusion in these regions.
To summarize, our multi-model analysis implies that the $\vec{B}$-field is a key factor in the energy dynamics of the CND. 
However, the $\vec{B}$-field may not be the primary interplaying-mechanism alongside the self-gravity of the gas and the tidal forces exerted by Sgr A*.

\subsubsection{Estimating Plasma \(\beta\): Assessing the Strength of $\vec{B}$ Field in the CND}
\label{sss:PlasmaBeta}

A straightforward assessment of the significance of the $\vec{B}$ field is to see Plasma Beta, $\beta_{\rm Plasma}$, value, which is given by a ratio of the gas and magnetic pressure as $\beta_{\rm Plasma} \equiv \frac{P_{\rm gas}}{P_{\rm mag}} = 2\left(\frac{c_{\rm iso}}{V_{\rm A}}\right)^2$. 
In the case of the CND, we calculate,
\begin{eqnarray}
&{\beta_{\rm Plasma}^{\rm na\acute{\i}ve}}& \simeq 0.2 \pm 0.1 \nonumber \\
& & \left(\frac{c_{\rm iso}}{0.63 \pm 0.48 ~\rm km ~s^{-1}}\right)^2
\left(\frac{V_{\rm A}}{2.2 \pm 0.9 ~\rm km~ s}^{-1}\right)^{-2}.
\label{eqn:PlasmaBeta}
\end{eqnarray}
Here, $c_{\rm iso}$ and $V_{\rm A}$ represent the isothermal sound speed and Alfv\'en velocity, respectively. 
We estimate the $c_{\rm iso}$ value, adopting a geometrical mean of the kinetic temperatures calculated from the LVG analysis of the multi-transition CS lines observations (200\,K; \cite{tsu18}) toward the CND and a representative value from the H$_2$CO observations (63\,K; \cite{gin16}).
The value of $V_{\rm A} = \frac{B_{\rm CND}^{\rm 3D}}{\sqrt{4\pi \rho}}$ is estimated using the Eq.(\ref{eqn:B_CND}) value and an estimate of ionized-and-neutral gas density $\rho$ (\S\ref{sss:DCF}), which is, practically, estimated from the number density of $(2.2\pm 1.4)\times 10^5\,\rm{cm^{-3}}$ from the LVG calculations (read from Figure 10a in \cite{tsu18}).
The variance of the Alfv\'en velocity, $(\Delta V_{\rm A})^2$, is estimated by $\left(\frac{\Delta B}{\sqrt{16\pi}\rho^{3/2}}\right)^2 +  \left(\frac{\Delta\rho}{\sqrt{4\pi\rho}}\right)^2$.
The obtained $\beta_{\rm Plasma}$ value is consistent with those reported in previous studies, such as \citet{Hsieh2018}, although their Eq.(5) omits the factor of 2.
A simplistic interpretation of Eq.(\ref{eqn:PlasmaBeta}) could suggest that the $\vec{B}$ field effectively confines the plasma, potentially affecting dynamics and heating of the gas (e.g., \cite{KAKIUCHI2024}).

Here, the inference of ${\beta_{\rm Plasma}^{\rm naive}} < 1.0$ does not reconcile with the results from the model analysis (\S\ref{sss:WKmodel}), which suggests that the $\vec{B}$-field pressure amplified via MRI dominates the gas pressure as the $\vec{B}$-field pressure saturates when it reaches up to the gas pressure. 
As discussed in \citet{Hsieh2018, gue23}, it is crucial to consider the turbulent pressure across the CND gas. 
Following the approach of Eq.(33) in \citet{gue23}, we estimate an effective Plasma $\beta$ as an upper limit of
\begin{eqnarray}
{\beta_{\rm Plasma}^{\rm eff}}
&\equiv& 
2\left(\frac{c_{\rm iso}^2 + \sigma_{\rm nth}^2}{V_{\rm A}^2}\right)
\lesssim \nonumber \\
& & 40 \left(\frac{\sigma_{\rm nth}}{10 ~\rm km ~s^{-1}}\right)^2
\left(\frac{V_{\rm A}}{2.2 \pm 0.9 ~\rm km ~s^{-1}}\right)^{-2}
\label{eqn:PlasmaBetaEff}
\end{eqnarray}
where we omit the small contribution of $c_{\rm iso}$.
The upper limit in Eq.(\ref{eqn:PlasmaBetaEff}) is based on the consideration that the observed $\sigma_{\rm nth}$ may not fully reflect turbulence contributions, as it could include line-width broadening due to systematic-gas motions such as infall and line-of-sight contaminations [see \S\ref{sss:DCF}, e.g., \cite{RSF14} as well]. 
Therefore, we selected the smallest estimate of $\sigma_{\rm nth} =$ 9.96 km s$^{-1}$ by \citet{hen16} among those considered (21 km s$^{-1}$ by \cite{tsu18}; 35 km ~s$^{-1}$ by \cite{gue23}), as it provides a conservative upper limit. 
Because ${\beta_{\rm Plasma}^{\rm eff}} \lesssim 40$ represents an upper limit, logically, we do not rule out the possibility of a ``strong" $\vec{B}$ field.
Nonetheless, given the observational challenges in estimating gas density, kinematic parameters, and $\vec{B}$-field information, along with the highly active nature of the CND, it would be prudent to consider ${\beta_{\rm Plasma}^{\rm eff}} \gtrsim 1$. 
Should this be the case, the theoretical condition for driving MRI cycles -- where the effective sound speed exceeds the Alfv\'enic velocity -- may be satisfied.
{
Finally, we describe a potential caveat regarding the use of the effective sound speed, which implicitly assumes an isotropic velocity field -- a condition that is unlikely to be met.
}

\subsubsection{Is MRI Working in the CND of the MWG?}
\label{sss:MRI}
To address the question posed at the end of the previous subsection, we examine whether the observed characteristics of the CND satisfy the fundamental criterion for MRI \citep{Vel59, cha61, BH91}: the wavelength of the most unstable mode  ($\lambda_{\rm max}$) must be smaller than the scale height of the CND, $H$, i.e., $\lambda_{\rm max} < H$.
The formula for $\lambda_{\rm max}$ is given by $\lambda_{\rm max} = \frac{8\pi}{\sqrt{15}} \frac{V_{\rm A}}{\Omega}$ where $\Omega$ represents the Keplerian angular velocity. 
The selection of $B_z$ is predicated on the assumption that the CND was initially threaded by a vertical $\vec{B}$ field, rendering the estimated $B_z$ from Eq.(\ref{eqn:B_z}) a plausible approximation. 
A Keplerian velocity $v_{\rm Kep} \simeq 90$ km s$^{-1}$, at a radius $r^{\rm 3D}_{\rm cav} \simeq 2.2$ pc for a central mass of $4.152\times 10^6 M_\odot$ \citep{EHT2022}, aligns closely with the observed rotational velocity $v_{\phi 0} = 105$ km s$^{-1}$. 
This agreement would support the Keplerian-rotation assumption.
Consequently, we calculate
\begin{eqnarray}
\lambda_{\rm max} &  = & \sqrt{\frac{16\pi^2}{15}} \frac{B_z}{\sqrt{\rho}} \frac{r}{v_{\phi0}} = 0.090^{+0.12}_{-0.10} \,{\rm pc} \\
& \simeq &  0.1\pm 0.1 \,{\rm pc} \nonumber \\
& &\left(\frac{B_z}{0.2\pm 0.05 ~{\rm mG}}\right) \left(\frac{\rho_{\rm tot}}{(1.0\pm 0.6)\times 10^{-18} {\rm \,g~cm^{-3}}}\right)^{-\frac{1}{2}}  \nonumber \\  
&\times& \left(\frac{r_{\rm cav}^{\rm 3D}}{2.2~{\rm pc}}\right) \left(\frac{v_{\phi0}}{105\pm 5\, {\rm km~s^{-1}}}\right)^{-1},
\label{eqn:lambda_max}
\end{eqnarray}
with the standard error in $\lambda_{\rm max}$ calculated from its variance via $\Sigma_i\left(\frac{\partial \lambda_{\rm max}}{\partial x_i}\right)^2(\Delta x_i)^2$ where $x_i$ encompasses $B_z$, $\rho_{\rm tot}$, $r_{\rm cav}^{\rm 3D}$, and $v_{\phi0}$, with their respective uncertainties detailed above. 
The uncertainty in $r_{\rm cav}^{\rm 3D}$ is assumed to be 20\%, and the contribution ratio of each term in the summation is 23:42:4.4:1, indicating that the uncertainty in $\rho_{\rm tot}$ predominantly determines the error.

We refine the analysis by focusing on the scale height estimation for the CND and its comparison with the $\lambda_{\rm max}$ parameter. 
The reported thickness of the CND ranges from 0.5 pc \citep{gus87} to 0.34 pc \citep{lau13}. 
Assuming these measurements denote the half-thickness, $z_{\rm half}$, we derive the relationship between the scale height, $2H$, and $z_{\rm half}$, represented as $H = \sqrt{\ln 2}/2 \cdot z_{\rm half} \simeq 0.42 \cdot z_{\rm half}$ for an exponential disk, and $1/\sqrt{2} \cdot z_{\rm half}$ for a Gaussian distribution. 
By averaging the thickness estimates, $z_{\rm half}$ is approximated to be $\sqrt{0.5 \cdot 0.34}$ pc, yielding $z_{\rm half} = 0.41 \pm 0.08$ pc, and leading to an observed scale height, $H_{\rm obs}$, of about $0.22 \pm 0.07$ pc for both disk models. 
This incorporates the error in $z_{\rm half}$ and the uncertainty in the conversion from $H$ to $z_{\rm half}$ due to the lack of direct observations of the CND's vertical structure.
Further, scale heights were estimated from the angular velocity, $\Omega = v_{\phi 0}/r_{\rm cav}^{\rm 3D} \approx (1.5 \pm 0.02)$ rad s$^{-1}$, and the isothermal sound speed, $c_{\rm iso}$, leading to $H_{\rm iso} = c_{\rm iso}/\Omega$. 
An analogous approach was used considering the effective sound speed, which includes the turbulence contribution, yielding $H_{\rm eff} = c_{\rm eff}/\Omega$. 
Consequently, the estimated scale heights and the probabilities that $H > \lambda_{\rm max}$ are:
\begin{itemize}
    \item $H_{\rm iso} \simeq 0.013 \pm 0.008$ pc, greater than $\lambda_{\rm max} \simeq 0.090^{+0.12}_{-0.10}$ pc (26\% probability),
    \item $H_{\rm eff} \simeq 0.21 \pm 0.12$ pc, surpassing $\lambda_{\rm max} \simeq 0.090^{+0.12}_{-0.10}$ pc (75\% probability), and
    \item $H_{\rm obs} \simeq 0.22 \pm 0.07$ pc, exceeding $\lambda_{\rm max} \simeq 0.090^{+0.12}_{-0.10}$ pc (82\% probability),
\end{itemize}
assuming both $\lambda_{\rm max}$ and $H$ follow normal distributions, with uncertainties represented as $1\sigma$ errors. 
Given the highly turbulent nature of the magnetized gas, disregarding the first scenario is reasonable. 
If the above analysis holds, asserting that $H$ should be greater than $\lambda_{\rm max}$ is statistically justifiable. 
However, further astrophysical clarification is required; attributing the high ($\vec{B}$ field environment of the CND solely to an ideal MRI is clearly insufficient.

We discuss the observed gas in the CND as being in a magnetically dominant and tidally stable state, rather than solely governed by gas pressure, as discussed in \citet{Hsieh2018}. 
The calculation of ${\beta_{\rm Plasma}^{\rm na\acute{\i}ve}} < 1$, which considers only the thermal pressure, alongside the low probability of $H_{\rm iso} > \lambda_{\rm max}$, supports the argument that MRI does not manifest in its idealized form due to the dominance of magnetic over gas pressure \citep{TKSuzuki2014}. 
Conversely, we propose that ${\beta_{\rm Plasma}^{\rm eff}} > 1$ more accurately reflects the condition of the gas, suggesting that the effective pressure of the gas exceeds the magnetic pressure. 
This inference is bolstered by the relatively high probabilities for $H_{\rm eff} > \lambda_{\rm max}$ and $H_{\rm obs} > \lambda_{\rm max}$, indicating a highly vigorous form of ``effective MRI," with a growth timescale of $2\pi/\Omega \sim 10^5$ years, resulting in the ``effective MRI" intermittent cycle of $\sim 10^6$ years, 10 times of the growth rate \citep{TKSuzuki2014}.
{
We conduct 
{a robust stability analysis} of the MRI interpretation, as detailed in Appendix \S\ref{a:Appendix}.
}

Given ${\beta_{\rm Plasma}^{\rm eff}}$ of the order of 10 and $H_{\rm obs} > \lambda_{\rm max}$, we note that the shear motions of the gas amplifies the $\vec{B}$ field. 
The stretching of field lines by gas motions results in increased Maxwell stress, as demonstrated by the numerical simulations given by  \citet{TKSuzuki2014} and  \citet{TKSuzuki2015}.
Angular momentum flux is conveyed by both Reynolds (velocity) and Maxwell (magnetic) stresses, defined as $\rho v_r \delta v_\phi - \frac{B_rB_\phi}{4\pi} \equiv \alpha_{\rm R} + \alpha_{\rm M}$ \citep{BH98}. 
From the multiplication of Eqs.~(\ref{eqn:B_phi}) and (\ref{eqn:B_r}), we estimate that $\alpha_{\rm M}$ would be in the order of $10^{-8}$ g cm$^{-1}$ s$^{-2}$ or less. 
This upper limit arises because the Maxwell stress must be determined by the correlation of $B_{\rm R}$ and $B_{\phi}$, which is generally smaller than their simple multiplication.
Currently, a reliable estimate for the $\alpha_{\rm R}$ term cannot be derived from existing molecular-line data.
Importantly, a well-ordered ($\vec{B}$ field is capable of transporting angular momentum outward, a process known as magnetic braking \citep{WD1967, TKSuzuki2014}.
This mechanism may challenge the traditional perspective of MRI's role in the CND, suggesting an intricate balance between magnetic and effective gas pressures in angular momentum transport and disk stability.

\section{Summary}
\label{s:Summary}

The study presented in this paper delves into the intricate gas dynamics of the CND surrounding Sgr A* at the Galactic Center, focusing on the effect of the $\vec{B}$-field  with a linear resolution of 0.5 pc. 
We present an $\lambda=$ 850 $\mu$m polarization map by utilizing archival data from two independent projects carried out at JCMT. We analyzed ancillary datasets taken with ALMA, VLA, and SOFIA to get a more comprehensive view. 
Our findings are summarized as follows.

\begin{enumerate}

\item The $P_{850}$ map shows a coherent-polarization structure across the CND with a well-defined radial-polarization profile, peaking towards Sgr A* and at $r\simeq 2.0$ pc. 
A visual comparison between the $P_{850}$ map and spectral-index map from \citet{gar11} yields that observed polarization at $r\gtrsim 0.5$ pc are attributed to the thermal emission from the grains. 
A comparison between the SOFIA's $\lambda=$ 37.1 $\mu$m map and the radial $P_{850}$ profile sets a characteristic scale with which we defined the size of the CND.

\item The CS $J=2-1$ emission comparison with the $\lambda=$ 850 and $\lambda=$ 37.1 $\mu$m continuum maps revealed two main findings: (i) the molecular gas associated with the CND extends beyond its boundaries, possibly linking it to the CMZ, and (ii) the emission at the center exhibits a less defined elliptical shape, likely affected by the 
SMBH's activity. 
Our centroid-velocity map revealed a complex structure, yet the overall rotational pattern of the CND was in agreement with previous studies. 
By applying a simple model to assess the CND's rotational velocity, we derived a value of 105 km s$^{-1}$, along with other geometrical parameters, aligning with previous studies.

\item The CND's $\vec{B}$-field structure exhibits a spiral pattern across a 1-pc scale centered around Sgr A*.
Comparative analysis using the \citet{war90} model found that the gas-pressure dominant models (gc1, gc2) and gas-and-magnetic-field comparable model, gc3, well reproduced the observations, particularly in regions corresponding or diverging from the CND's ring. 
The morphology and the models' assumptions indicate a predominantly toroidal $\vec{B}$ field, shaped by accretion dynamics, with the observed field geometry playing a crucial role in the CND's energetics and dynamics.

\item Utilizing residuals from the \citet{war90} model, we applied the DCF method to estimate the plane-of-sky $\vec{B}$-field strength of the CND, resulting in upper limits of the order of 10 mG. 
We discussed possible causes of this failure.

\item\label{summary:MRIpossibility} Based on the modeling, we quantified the magnetic-field strength at the inner-ionized cavity's center surrounding Sgr A* and within the Circumnuclear Disk (CND) using the enhanced sensitivity of the POL-2 system at JCMT. 
Our analysis revised the \(\alpha_{\rm WK}\) and \(\beta_{\rm WK}\) parameters, leading to an estimated \(|B_0^{\rm 3D}|\) $\vec{B}$-field strength of approximately $0.24^{+0.05}_{-0.04} 0.1$ mG. 
Further analysis suggests a predominantly toroidal $\vec{B}$-field structure in the CND, with \(B_r\), \(B_\phi\), and \(B_z\) components derived as $0.4 \pm 0.1$ mG, $-0.7 \pm 0.2$ mG, and $0.2 \pm 0.05$ mG, respectively. 
A critical finding is the dominance of \(|B_\phi|\) over \(B_r\) and \(B_z\), indicating that the \(\vec{B}\)-field component, which is amplified by differential rotation, is considered to be prominent.

\item Motivated by the ``pure MRI" interpretation in the item N.\ref{summary:MRIpossibility}, we calculate the Plasma Beta.
The naive estimation of ${\beta_{\rm Plasma}^{\rm na\acute{i}ve}} \simeq 0.2 \pm 0.1$ implies that magnetic pressure dominates over gas pressure. 
However, when taking turbulent pressure into account, we obtained an effective \(\beta_{\rm Plasma}^{\rm eff} \lesssim 40\), suggesting that the ``effective gas pressure," which includes the turbulent pressure, would exceed the magnetic pressure in the CND.
Careful interpretation is required when considering ``idealized MRI cycles", especially if the Alfv\'enic velocity, $2.2 \pm 0.9$ km s$^{-1}$, exceeds the isothermal sound speed of $0.63 \pm 0.48$ km s$^{-1}$.

\item Assessing the potential for MRI cycles in the CND involves comparing $\lambda_{\rm max}$, the wavelength associated with MRI's fastest growth rate, against the CND's scale height, $H$. 
For MRI to occur, the condition $\lambda_{\rm max} < H$ must be satisfied. 
Our calculations suggest the CND might meet the necessary conditions for MRI, particularly when considering the effective sound speed to reflect the turbulent nature of the CND. 
Three estimates of scale heights, $H_{\rm iso}$, $H_{\rm eff}$, and $H_{\rm obs}$, when compared to $\lambda_{\rm max}$, imply the effective gas pressure could exceed magnetic pressure, allowing for ``effective MRI" with relatively ordered $\vec{B}$ field transports angular momentum. 
Nonetheless, the inherent complexities in measuring key CND parameters underscore the need for cautious interpretation and further research to fully elucidate MRI's role in the CND's dynamics.

\end{enumerate}

The presented study demonstrated the effectiveness of linear polarization imaging in elucidating the gas dynamics within the CND, highlighting the interplay between the $\vec{B}$ field, turbulence, and gravity. 
The next challenge lies in achieving high-dynamic range polarimetric imaging -- a high ratio of image size to resolution -- to disentangle the $\vec{B}$-field structure. 
Equally imperative is the development of innovative techniques to resolve the complexities posed by line-of-sight confusion in molecular-line emissions, a challenge that, once overcome, will significantly deepen our understanding of the intricate structures and processes governing the CND.

\begin{ack}
The authors thank Dr. Tomisaka for a fruitful discussion. This research used the Astropy Suite (Astropy Collaboration 2013, 2018). The James Clerk Maxwell Telescope is operated by the East Asian Observatory on behalf of The National Astronomical Observatory of Japan; Academia Sinica Institute of Astronomy and Astrophysics; the Korea Astronomy and Space Science Institute; the National Astronomical Research  nstitute of Thailand; Center for Astronomical Mega-Science (as well as the National Key R$\&$D Program  f China with No. 2017 YFA0402700). Additional funding support is provided by the Science and Technology Facilities Council of the United Kingdom and participating universities and organizations in the United Kingdom and Canada. Additional funds for the construction of SCUBA-2 were provided by the Canada Foundation for Innovation. This paper makes use of the fol owing ALMA data: ADS/JAO.ALMA\#2017.1.00040.S. ALMA is a partnership of ESO (representing its member states), NSF (USA), and NINS (Japan), together with NRC (Canada), MOST and ASIAA (Taiwan), and KASI (Republic of Korea), in cooperation wi h the Republic of Chile. The Joint ALMA Observatory is operated by ES  AUI/NRAO, and NAOJ. This work is supported in part by a Grant-in-Aid for Scientific Research of Japan (17H01105; 17K05388; 19H01938; 21H00033; 22H01263) and the MEXT program supporting female and young researchers in research projects at Kagoshima University. This research made use of Astropy (Astropy Collaboration 2013, 2018).  Data analysis was in part carried out on the Multi-wavelength Data Analysis System operated by the Astronomy Data Center (ADC), National Astronomical Observatory of Japan. 
\end{ack}

\appendix
\section{A Robust Stability Analysis of the MRI Interpretation and Its Implications}
\label{a:Appendix}
~~~~~Here we briefly assess the MRI interpretation within the instability parameter space, as conducted by \citet{Hopkins2023_Bfield} in their cosmological radiation-magnetohydrodynamic (RMHD) simulations addressing the role of $\vec{B}$ fields around the SMBH.
\citet{Hopkins2023_Overall} developed RMHD simulations to investigate the formation and evolution of a quasar (QSO) accretion disk. 
Specifically, \citet{Hopkins2023_Bfield} furthered the analytical calculations of \citet{Pessah2005} to examine the toroidal stability parameter of the disk, $Q_\phi^{\rm eff} = \frac{v_{\rm A, \phi}}{c_{\rm iso}}$, and its height, $Q_z^{\rm eff} = \frac{2\pi r}{H}\frac{v_{\rm A, z}}{c_{\rm iso}}$, in the vertical direction. 
Here, $v_{\rm A, \phi}$ and $v_{\rm A, z}$ represent the azimuthal and vertical components of the Alfvén speed, respectively. 
From our observations, we derive the following stability parameters:
\begin{equation}
{Q_\phi^{\rm eff}}\simeq
0.2^{+0.5}_{-0.2} 
\left(\frac{|V_{\rm A,\phi}|}{1.8\pm 0.77 ~\rm km ~s^{-1}}\right)
\left(\frac{c_{\rm eff}}{10~\rm km ~s^{-1}}\right)^{-1},
\label{eqn:Q_phi}
\end{equation}
and 
\begin{eqnarray}
{Q_z^{\rm eff}} &\simeq&
4^{+5}_{-4} 
\left(\frac{r_{\rm cav}^{\rm 3D}}{2.2~\rm pc}\right)
\left(\frac{H_{\rm eff,obs}}{0.22 \pm 0.10~\rm pc}\right)^{-1} \nonumber \\
& &  \left(\frac{|V_{{\rm A,} z}|}{0.66 \pm 0.26 ~\rm km ~s}^{-1}\right)
\left(\frac{c_{\rm eff}}{10~\rm km ~s^{-1}}\right)^{-1}.
\label{eqn:Q_z}
\end{eqnarray}
{All the superscripts and subscripts} of 
``eff'' in the \(Q\) calculations are due to the adoption of the effective sound speed (\S\ref{sss:PlasmaBeta}), and we incorporated the values of \(H_{\rm eff}\) and \(H_{\rm obs}\) in the calculations of \(Q_z^{\rm eff}\).
When comparing the derived $Q_\phi^{\rm eff}$ and $Q_z^{\rm eff}$ values with the lower right panel of Figure 16 of \citet{Hopkins2023_Bfield} and Figure 3 of \citet{Pessah2005}, the instability we observe for the CND is clearly within the MRI region. 
The instability parameters are notably situated within the MRI region [Region I in \citet{Pessah2005}], reinforcing our hypothesis that MRI predominates over the CND.

\begin{figure}[ht]
\includegraphics[width=8.2cm]{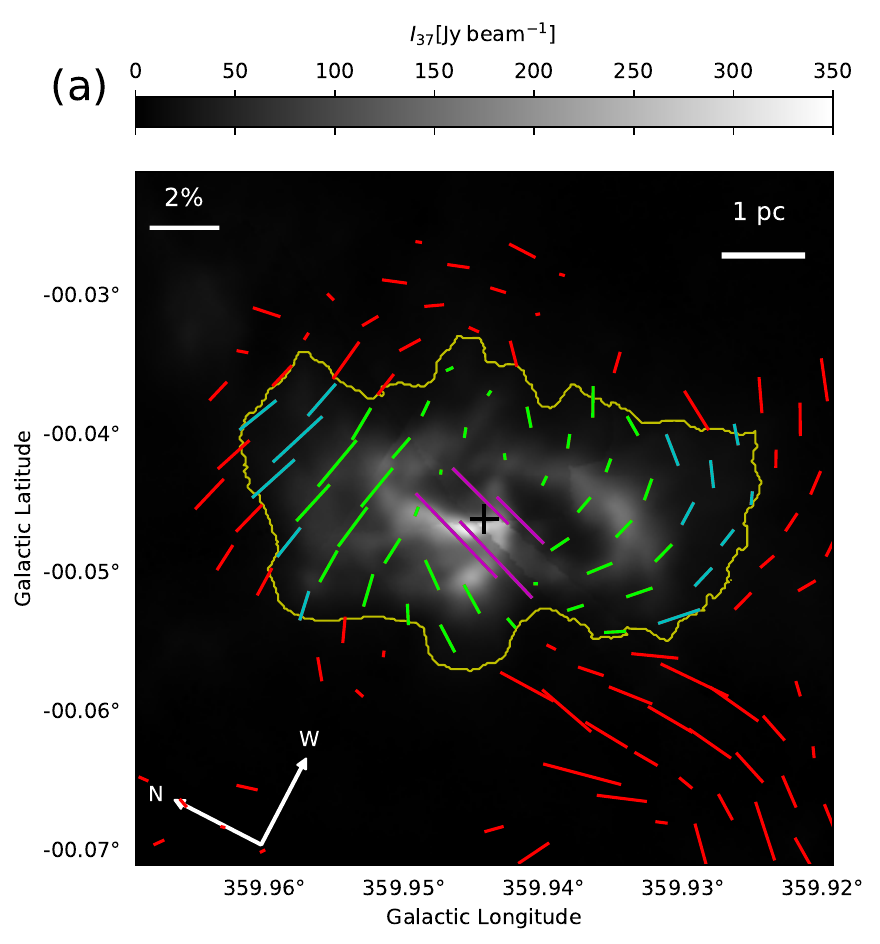}
\includegraphics[width=8.2cm]{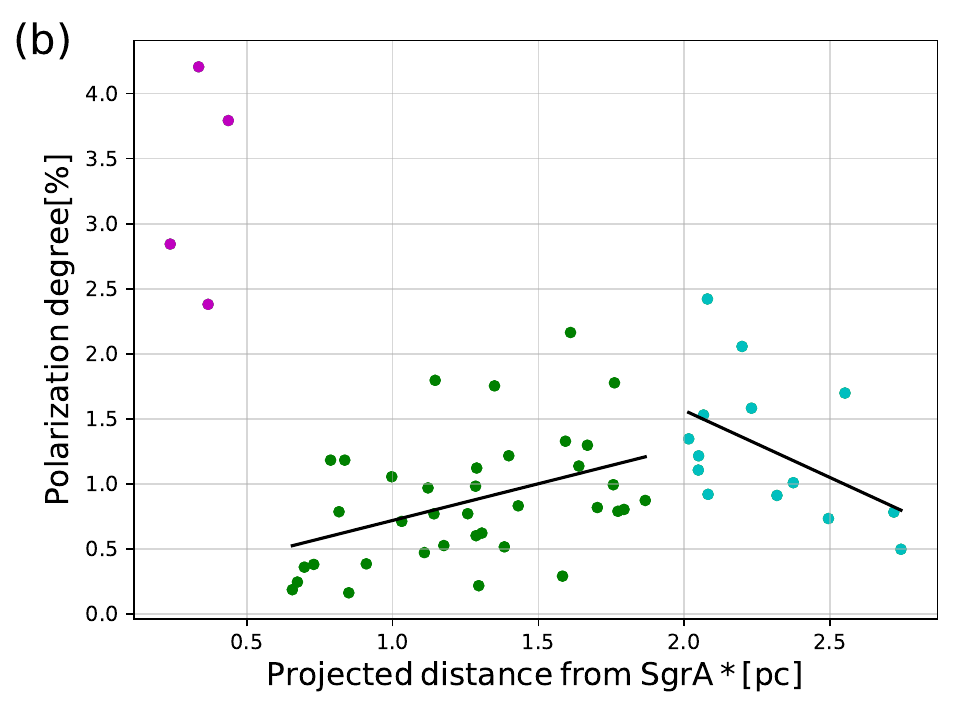}
\caption{
(a) $\lambda=$ 850 $\mu$m polarization map overlaid on the $\lambda=$ 37.1 $\mu$m continuum image. 
The lengths of the color segments are proportional to the $P_{850}$ values, scaled with the horizontal-white bar at the top-left corner, and the directions of the segments represent the polarization angles at each position.
All the segments are shown with a 12\farcs0 grid-spacing, comparable to the beam size (12\farcs6).
Here, we display only those segments that exhibit a signal-to-noise ($S/N$) ratio in the Stokes $I$ greater than 50. It is important to note that we did not perform any data clipping based on the $S/N$ of $P$ values, in order to avoid missing intrinsically weakly polarized emissions.
The green segments represent polarization detected within a circle of radius $R_{\rm eff} = 2.0$\,pc. The cyan segments are found inside the $20\sigma$-level contour of the $\lambda=$ $37.1 \mu$m continuum emission (as indicated by the yellow contour), but at projected distances $r_{\rm pos} \geq 2.0$\,pc. The red segments are located outside this region.
The central cross indicates the position of Sgr A*. 
(b) A scatter plot of the $P_{850}$ values as a function of the projected distance in a unit of pc from the Sgr A*.
All the data points are taken from panel (a), and the colors in the scatter plot correspond to those in (a).
Notice that the error bars associated with the $P_{850}$ values are smaller than the symbols.
The blue and red lines are the best fits for the projected-distance ranges between $0.5$ pc and $2.0$ pc, and $\geq 2.0$ pc, respectively. 
See Table \ref{tbl:POL2Results} for the values used to make the plots.
\label{fig:PinCNDaround}
}
\end{figure}

\begin{figure}[ht]
\includegraphics[width=8.2cm]{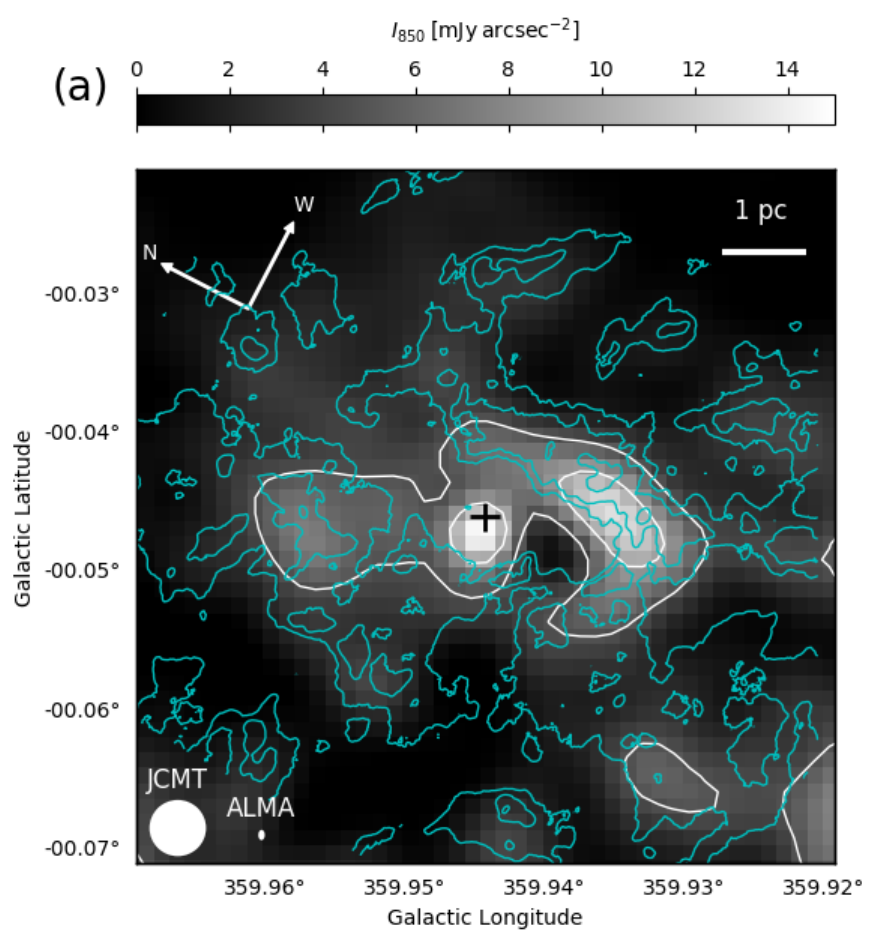}
\includegraphics[width=8.2cm]{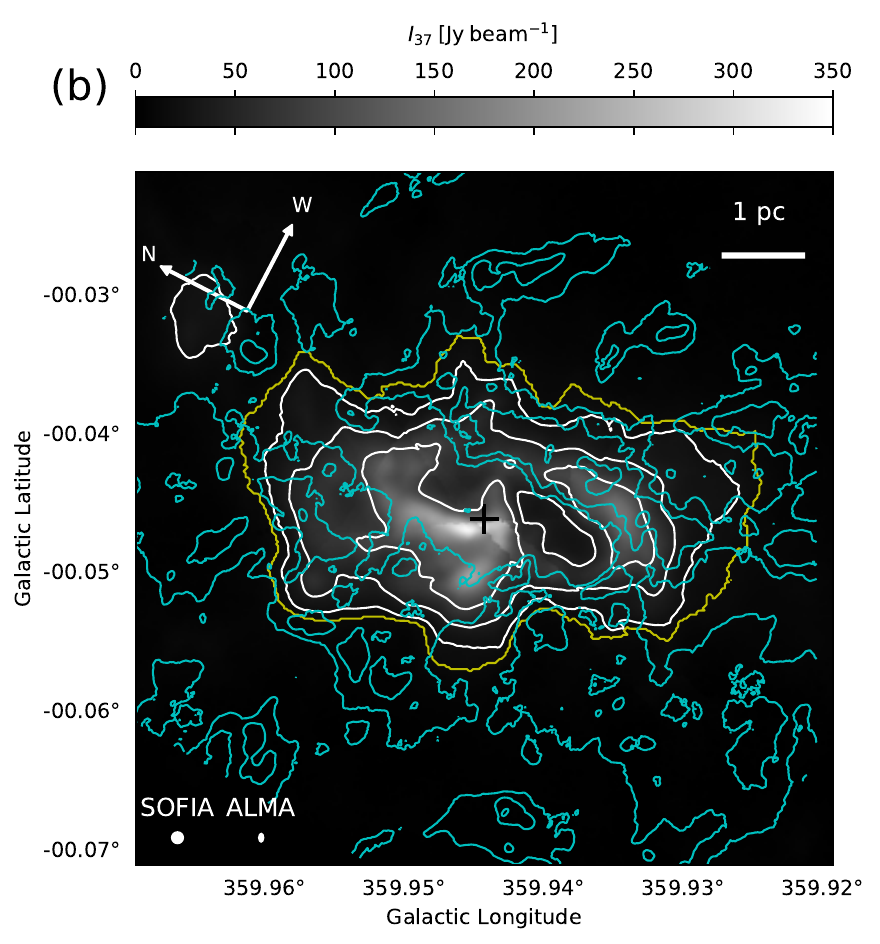}
\caption{
Comparisons of the overall distribution of the CS $J=2-1$ emission (blue contours) with (a) the $\lambda=$ 850 $\mu$m and (b) $\lambda=$ 37.1 $\mu$m continuum emission (white contours and gray-scale images). 
The CS $J=2-1$ emission map is an integrated intensity map of the CS $J=2-1$ emission obtained as a zeroth-order momentum map calculated over an LSR-velocity range of $-150 \leq v_{\mathrm{LSR}}/\mathrm{[km\,s^{-1}]} \leq 150$ with a detection threshold at the $3\sigma$ level of each channel.
The contour intervals and others are drawn in the same fashion adopted in Figure \ref{fig:ContImages}.
Here, the RMS noise level of the CS map is 0.78\,mJy\,beam$^{-1}\,\cdot\mathrm{km\,s^{-1}}$.
\label{fig:CS21on850and37}
}
\end{figure}

\begin{figure*}[ht]
\includegraphics[width=18cm]{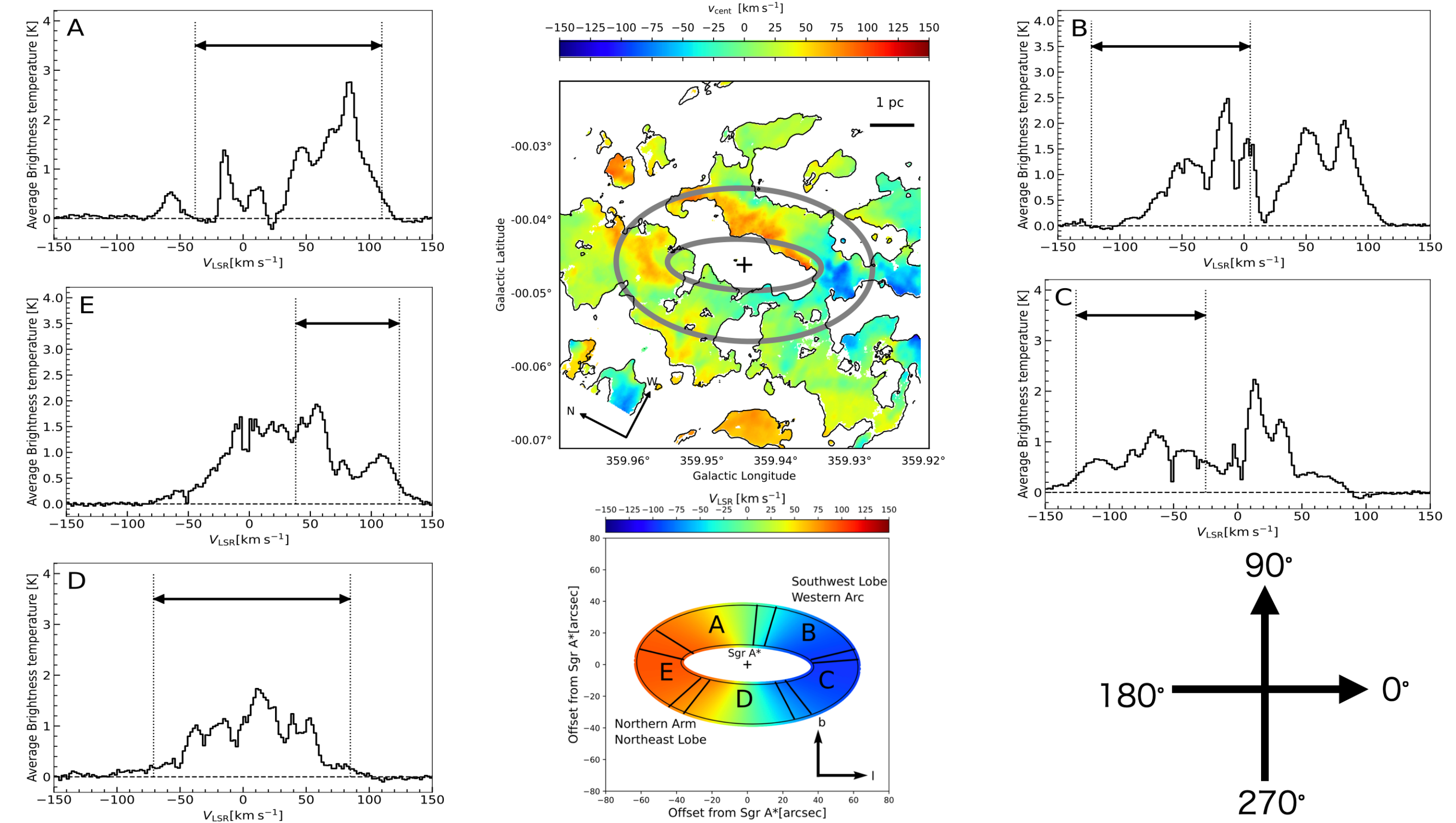}
\caption{
The upper-middle panel presents an overlay of the total-integrated intensity map of the CS $J=2-1$ emission (contours) on the velocity-centroid map of the line (color).
The contour represents the 5$\sigma$ level in the same fashion as those in Figure \ref{fig:CS21on850and37}.
The two thick gray ellipses show the location and extent of the modeled ring in \S\ref{sss:VF}.
The lower-middle panel displays the line-of-sight velocity of the modeled circumnuclear disk (CND) ring. 
Alphabetical labels on the ring correspond to the names of the subregions, defined by dual-radial segments on their sides. Refer to Table \ref{tbl:Regions} for the azimuthal angles, $\phi$, used in Eq.(\ref{eqn:Vrot}) to define these dual-radial segments. 
The labels, such as the Northern Arm, are designated for the known structures associated with each subregion.
The two orthogonal arrows in the lower-right corner show the $\phi$ coordinate defined in the model for the convenience of comparisons.
The surrounding panels show the spectral lines of the CS $J=2-1$ emission observed by ALMA, produced by integrating the emission over the spatial extent of each subregion; the vertical axes of the spectra represent the mean beam-averaged brightness temperature over the individual components identified by the ring model.
\label{fig:CS21spGuideMap}
}
\end{figure*}

\begin{figure}[ht]
\begin{center}
\includegraphics[width=8.2cm]{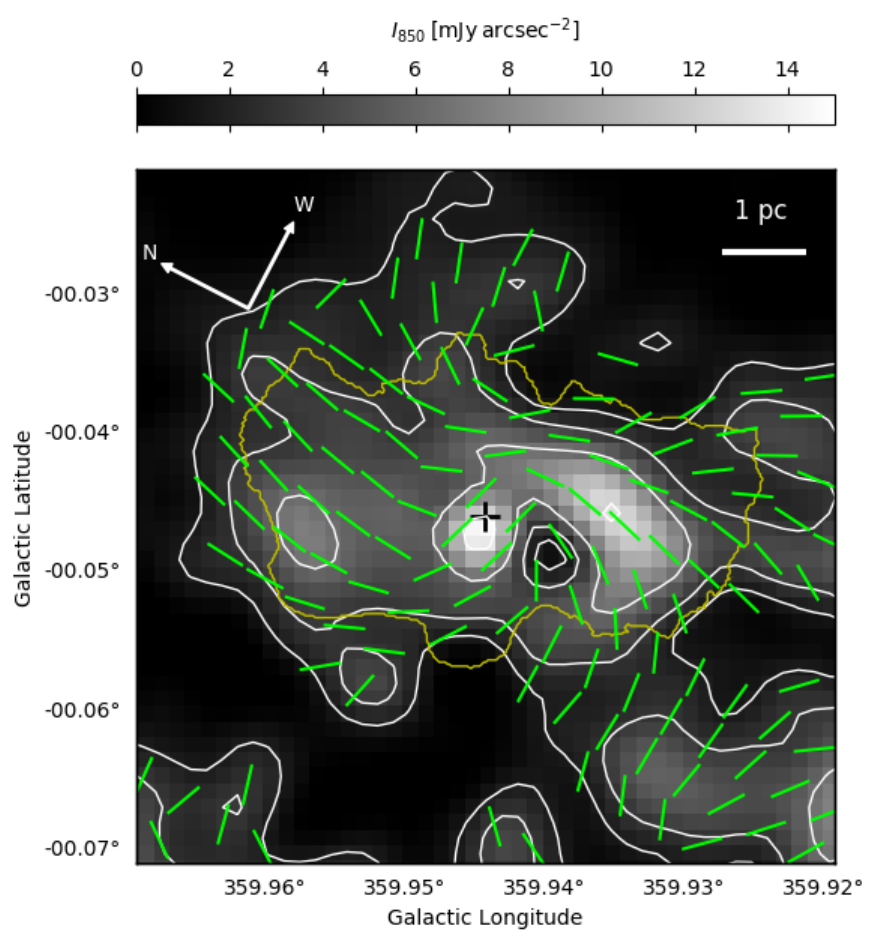}
 \end{center}
\caption{
Overlays of the inferred $\vec{B}$-field  direction (green segments of identical length) on the total intensity, that is, the Stokes \(I\) map (
{white  contours}), as presented by the grayscale image in Figure \ref{fig:ContImages}.
The intervals of the white contours are drawn in the same manner as those in Figure \ref{fig:ContImages}.
The yellow contour represents the 20$\sigma$-level contour of the 37 $\mu$m emission.
The identical-length green segments indicate \(\vec{B}\)  field directions which are determined by polarization direction rotated by 90 degrees. 
}
\label{fig:Bmaps}
\end{figure}

\begin{figure*}[ht]
\begin{center}
\includegraphics[width=6cm]{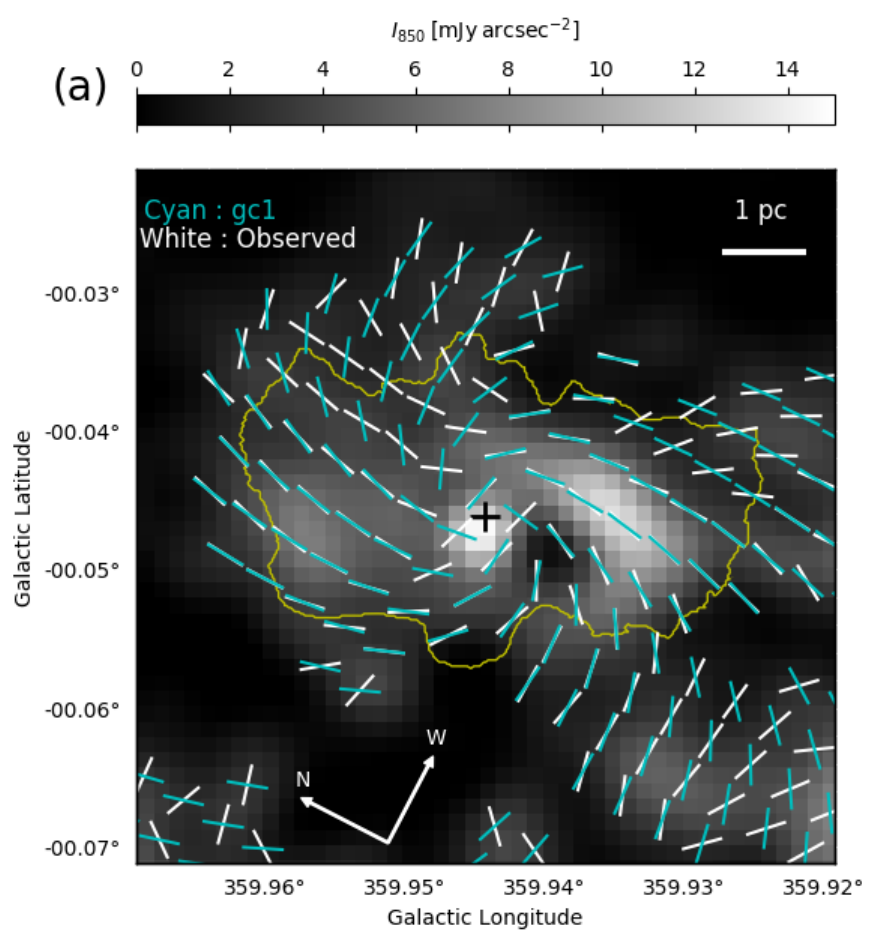}\includegraphics[width=6cm]{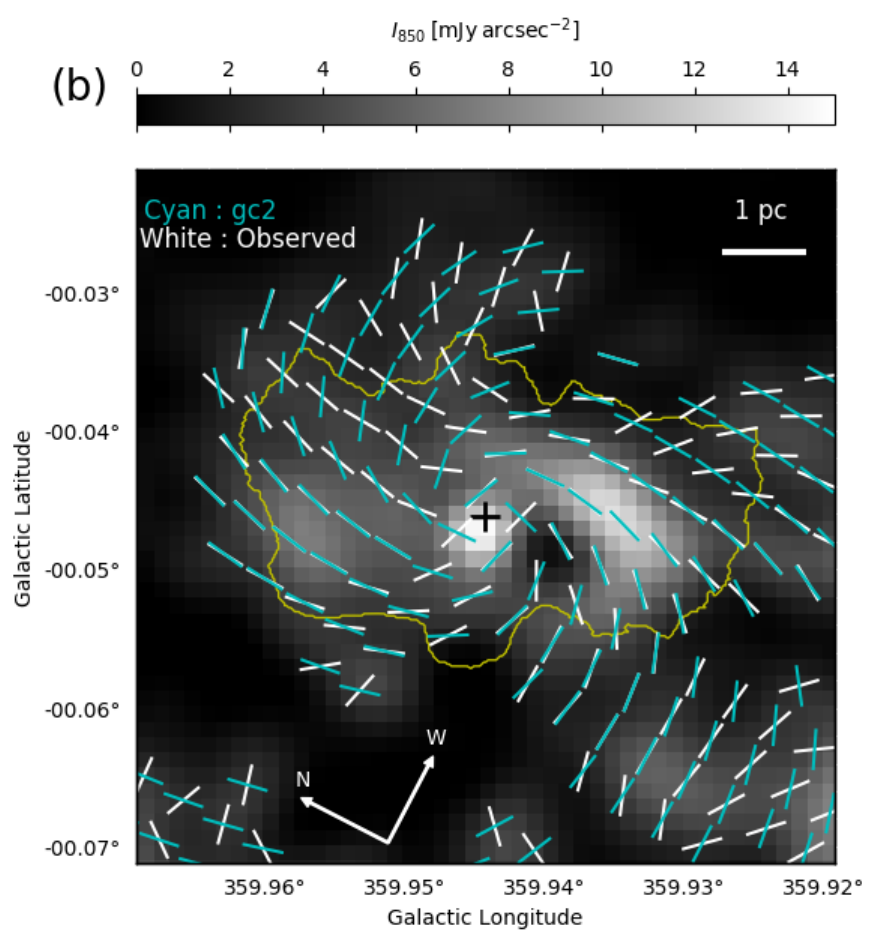}\includegraphics[width=6cm]
{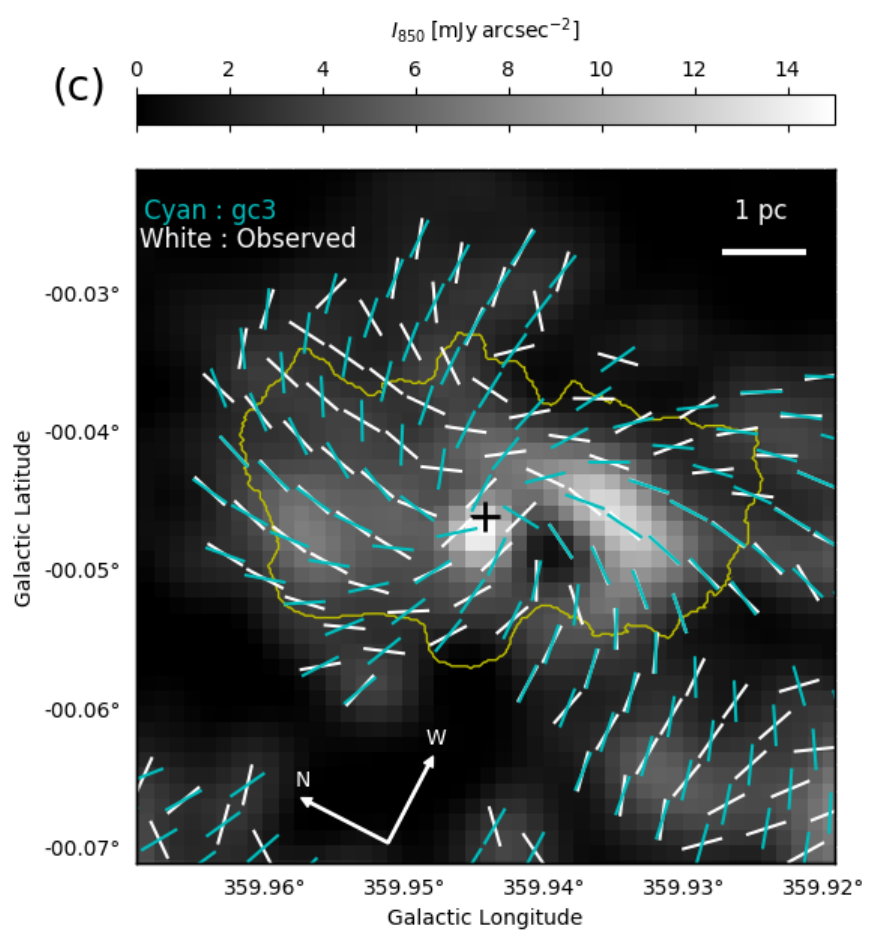}
 \end{center}
\caption{
Panels (a), (b), and (c) display comparisons of the inferred $\vec{B}$ field (white segments; see Figure \ref{fig:Bmaps}) with predictions from the gc1, gc2, and gc3 models of \citet{war90}, respectively. 
For the resultant \(\chi^2\) values of each fitting, see Table \ref{tbl:model_fits}.
The grayscale image in each panel depicts the $\lambda=$ 850 $\mu$m continuum emission. 
The yellow contour represents the 20$\sigma$-level contour of the 37 $\mu$m emission, which is used to define the CND. 
Note that applying the model in regions far from the CND may not be appropriate.
}
\label{fig:GC123models}
\end{figure*}

\begin{figure*}[ht]
\begin{center}
\includegraphics[width=6cm]{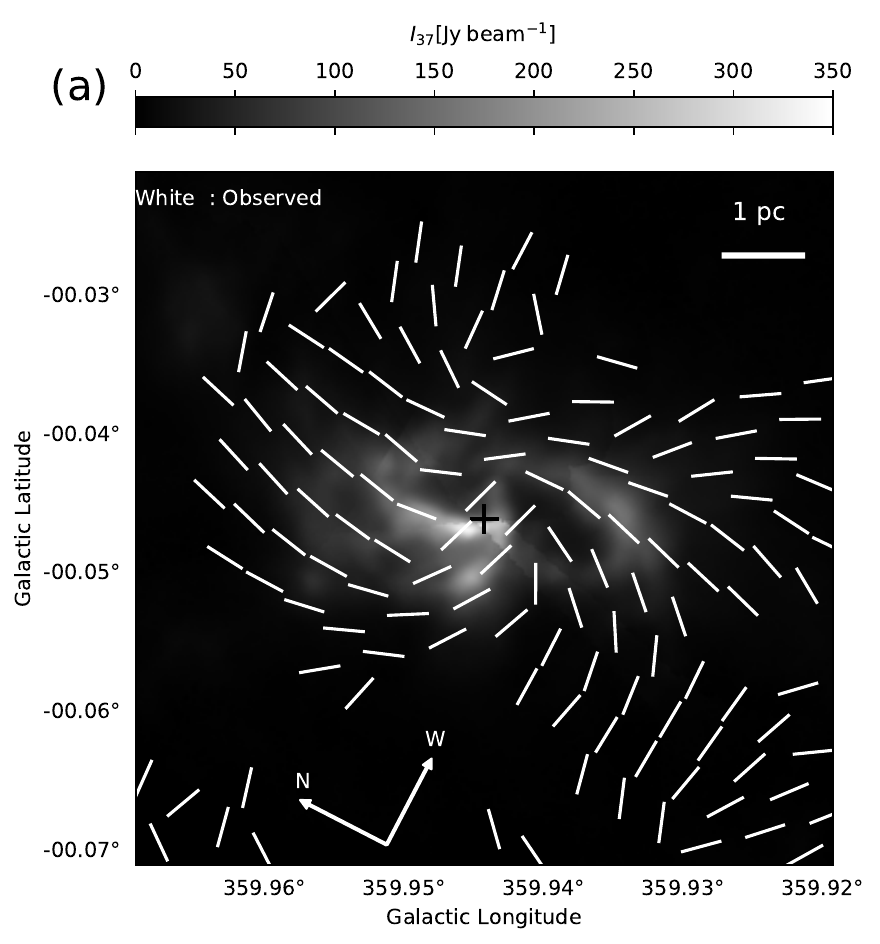}\includegraphics[width=6cm]{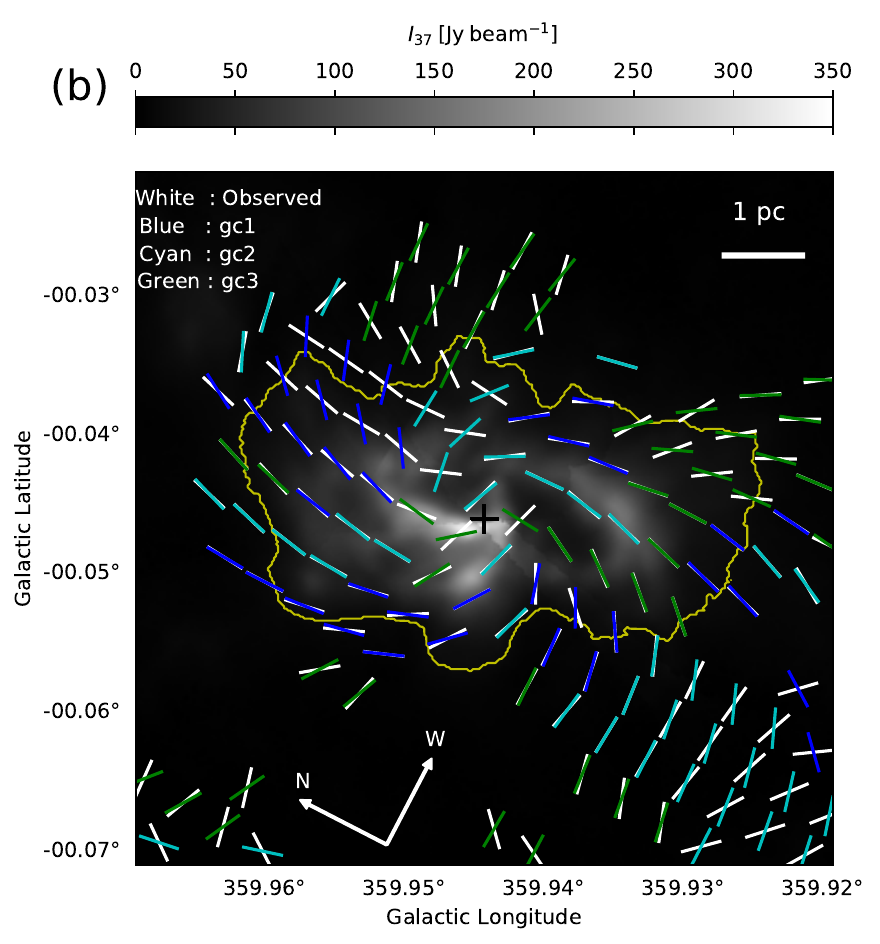}\includegraphics[width=6cm]{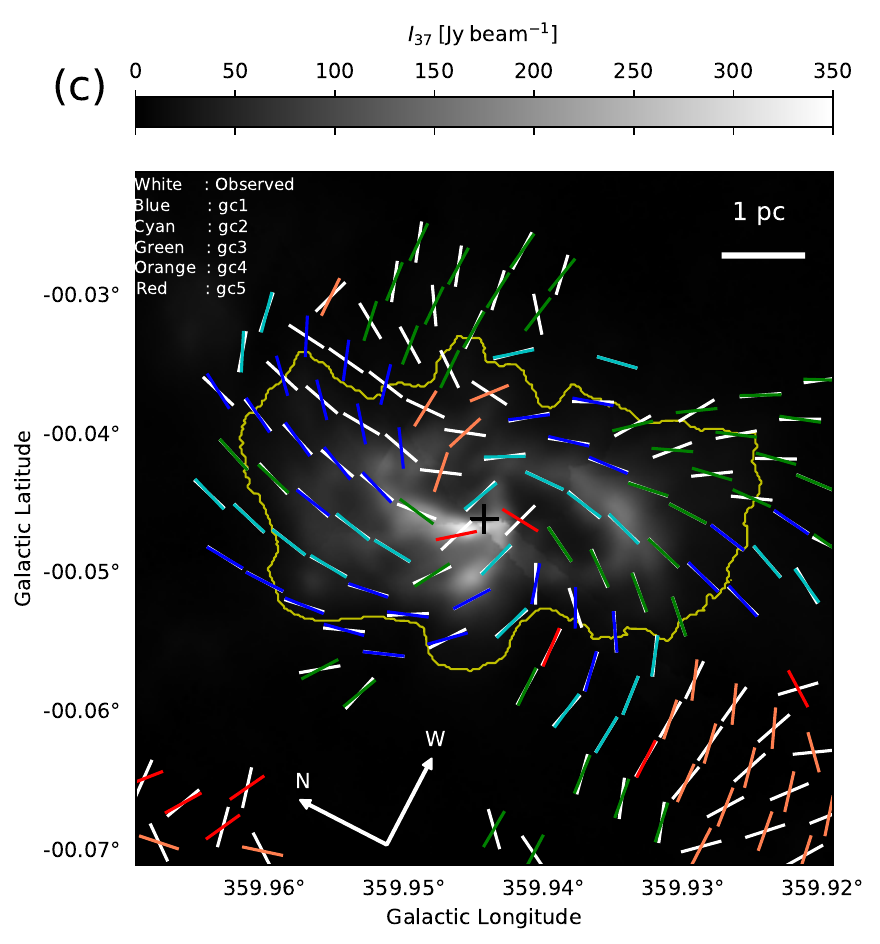}
 \end{center}
\caption{
The diagram (a) shows the spiral  $\vec{B}$-field directions in white segments overlaid on the $\lambda=$ 37.1 $\mu$m continuum emission image (gray scale).  The diagrams (b) and (c) show the comparisons of the $\vec{B}$-field directions between the observed (white segments) and predicted directions based on the different WK models, noted on each diagram.   
The notes in the upper-left corner explain the correspondences between the adopted models and their 
colors. 
The yellow contour represents the $20\sigma$-level contour, as shown in Figure \ref{fig:ContImages}b.
}
\label{fig:GC12345models}
\end{figure*}

\begin{figure}[ht]
\begin{center}
\includegraphics[width=7.5cm]{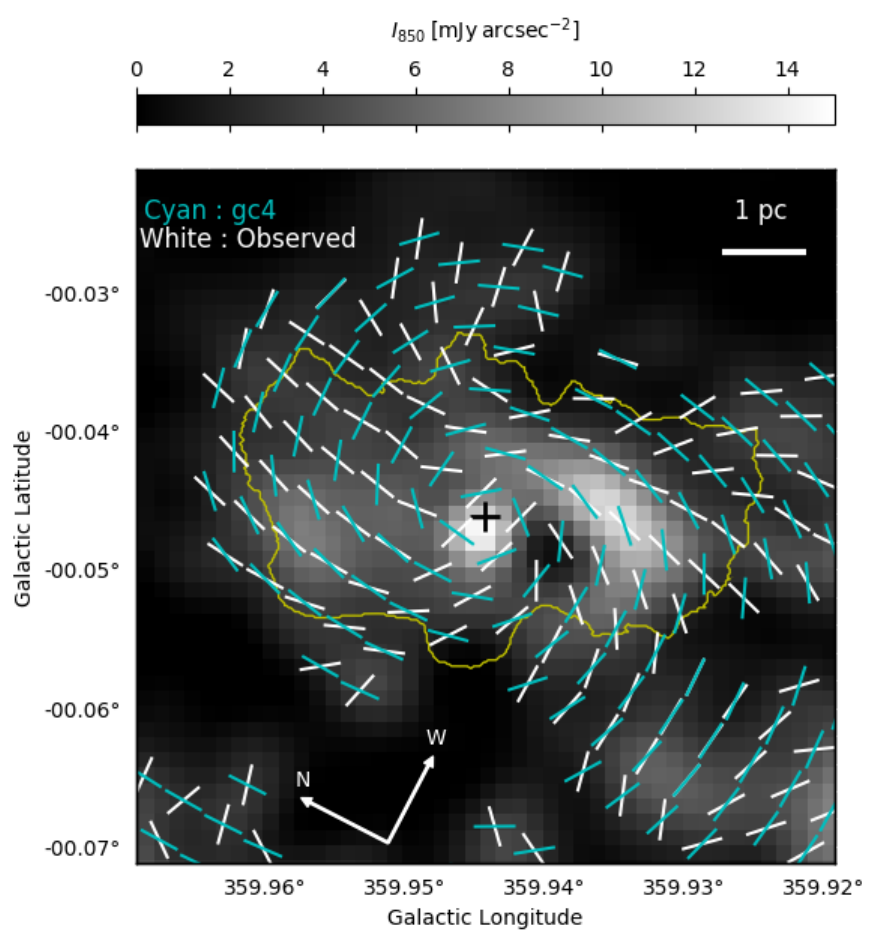}
\includegraphics[width=7.5cm]{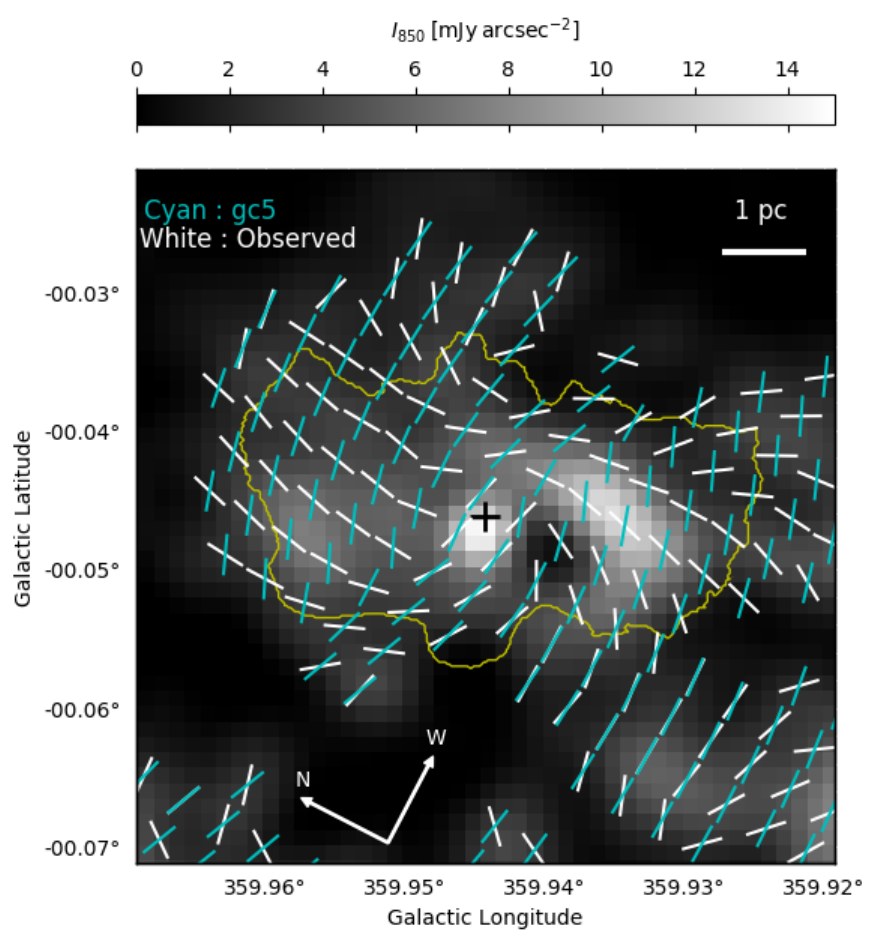}
 \end{center}
\caption{
Comparisons of the gc4 (
{upper}) and gc5 (
{lower}) models from \citet{war90} (depicted with cyan segments) with the inferred $\vec{B}$ field (represented by white segments), as shown in Figure \ref{fig:Bmaps}. 
The grayscale image and contours are consistent with those presented in Figure \ref{fig:GC123models}. 
For the resultant \(\chi^2\) values, see Table \ref{tbl:model_fits}.
}
\label{fig:gc4gc5}
\end{figure}

Currently, the CND is considered to be in a dynamically stable state, as deduced from the estimation of the \(r\sim 3\,\rm pc\)-averaged Toomre's \(Q\) parameter of 4--10 \citep{Oka2011}.
Conversely, the magnetic $Q_{\rm{Mag}}^{\rm{gas}}$ [refer to Figure 12 in \citet{Hopkins2023_Overall}] at a distance of 1\, pc from an SMBH may vary from 2 down to unity.
We, therefore, speculate that the MRI-active CND could be in a dynamically critical state at the scale we observed (see Figure~\ref{fig:PinCNDaround}b).

We need to be careful in the context of the prior discussion.  
The mass accretion rate we derived at approximately 2.2 pc from the SMBH is on the order of $10^{-2}$ $M_\odot$ yr$^{-1}$, as shown in Eq. (\ref{eqn:dMdt}). 
This distance from Sgr A* corresponds to the ``{\it Torus $\rightarrow$ BHROI}" (Black Hole Radius of Influence) region as outlined by \citet{Hopkins2023_Overall}, where the predicted mass accretion rate is approximately $10^2$ $M_\odot$ yr$^{-1}$, significantly higher by four orders of magnitude than our estimate. 
Although the accretion rate in the inner $\sim$0.1--1 pc ``{\it Torus $\rightarrow$ Non-Star-Forming Disk}" region delineated by \citet{Hopkins2023_Overall} decreases slightly (refer to their Figure 7), it remains on the order of 10 $M_\odot$ yr$^{-1}$ near the innermost stable circular orbit (ISCO) for a $5\times 10^7$ $M_\odot$ black hole in the QSO-accretion-disk model. 
Therefore, the application of the Hopkins model to our Galactic CND requires careful consideration, emphasizing the need for improved observational techniques and model refinement.
Further details will be described elsewhere.

\clearpage
%
%

%
%
%
%
\clearpage
\begin{longtable}[!t]{lcccc}
\caption{
Datasets Used for This Study --- All the data were retrieved from the archive. 
The ALMA images reproduced by us were obtained by combining visibility data from three arrays: the 12-m arrays (14-minute exposure, yielding 0\farcs74$\times$0\farcs40 beam, and a 7-min 3\farcs35$\times$2\farcs11 beam), the 7-m array (82-minute exposure, yielding 18\farcs27$\times$8\farcs43 beam), and the Total Power array (162-minutes with 66\farcs0 beam). 
For the Bands/Lines column, we summarized the wavelength for the continuum emission and the transition for the emission line observations.
For image sensitivities, refer to the respective figure captions.
}
\label{tbl:Obs}
\hline
Telescopes/Arrays & Bands/Lines & Beam size & Integrated time (h) & Project or Proposal ID\\
\hline
\endhead
\hline
\endfoot
\hline
\endlastfoot
JCMT POL-2 & $\lambda = 850\,\mu$m & 12\farcs6  & 14.6 & M16BP061 and M17AP074 \\
VLA  &  $\lambda =$ 6\,cm & 9\farcs56$\times$3\farcs79 & 1.82 & AZ44\\
SOFIA/FORCAST & $\lambda = 37.1\,\mathrm{\mu m}$ & 3\farcs73 & 0.43 & 07-0189 \\
ALMA & CS $J=2-1$  & 2\farcs67$\times$1.67 
& 4.42 & 2017.1.00040.S \\
\hline
\end{longtable}

\clearpage
\begin{longtable}{lllccc}
\caption{
Results of the $\lambda=$ 850-\textmu m Polarization Measurement --- 
$l$ and $b$ denote the Galactic Longitude and Latitude, respectively, $r_{\rm pos}$ represents the projected distance from Sgr A* in units of pc, Color corresponds to those in Figure \ref{fig:PinCNDaround}, $P_{850}$ refers to the polarization degrees at the $\lambda=$ 850 $\mu$m band, and $\chi$ represents the polarization angles.
}
\label{tbl:POL2Results}
\hline
\multicolumn{1}{c}{$l$} & 
\multicolumn{1}{c}{$b$} & 
\multicolumn{1}{c}{$r_{\rm pos}$} & 
\multicolumn{1}{c}{\raisebox{-1.5ex}{Color}} & 
\multicolumn{1}{c}{$P_{850}$} & 
\multicolumn{1}{c}{$\chi$} \\
(degrees) & (degrees) & (pc) & & (\%) & (degrees) \\ 
\endfirsthead
\hline
\endhead
\endfoot
\hline \hline
\endlastfoot
\hline
359.9213 & $-0.07043$ & 4.8 & Red & 1.4$\pm$0.07 & $-33.4\pm 1.35$ \\
359.9195 & $-0.06758$ & 4.7 & Red & 0.9$\pm$0.05 & $-40.9\pm 1.54$ \\
359.9241 & $-0.06869$ & 4.3 & Red & 2.2$\pm$0.08 & $-44.6\pm 1.05$ \\
359.9224 & $-0.06584$ & 4.2 & Red & 1.2$\pm$0.06 & $-39.6\pm 1.50$ \\
359.9207 & $-0.06300$ & 4.1 & Red & 0.4$\pm$0.07 & $-56.9\pm 5.78$ \\
359.9287 & $-0.06980$ & 4.0 & Red & 1.7$\pm$0.11 & $-47.4\pm 1.60$ \\
359.9270 & $-0.06695$ & 3.9 & Red & 1.1$\pm$0.08 & $-34.2\pm 1.97$ \\
359.9252 & $-0.06411$ & 3.7 & Red & 1.4$\pm$0.09 & $-20.6\pm 1.70$ \\
359.9235 & $-0.06126$ & 3.7 & Red & 1.2$\pm$0.09 & $-21.4\pm 2.36$ \\
359.9218 & $-0.05842$ & 3.7 & Red & 0.6$\pm$0.17 & $-46.2\pm 7.47$ \\
359.9316 & $-0.06806$ & 3.6 & Red & 0.4$\pm$0.08 & $19.9\pm 5.67$ \\
359.9298 & $-0.06522$ & 3.4 & Red & 0.6$\pm$0.06 & $-12.3\pm 3.29$ \\
359.9281 & $-0.06237$ & 3.3 & Red & 1.8$\pm$0.13 & $-7.7\pm 1.46$ \\
359.9263 & $-0.05953$ & 3.2 & Red & 2.1$\pm$0.20 & $-8.1\pm 2.36$ \\
359.9211 & $-0.05099$ & 3.4 & Red & 0.7$\pm$0.07 & $57.9\pm 3.50$ \\
359.9194 & $-0.04814$ & 3.6 & Red & 0.8$\pm$0.07 & $-88.6\pm 2.41$ \\
359.9344 & $-0.06633$ & 3.2 & Red & 1.8$\pm$0.09 & $20.2\pm 1.47$ \\
359.9327 & $-0.06348$ & 3.0 & Red & 1.0$\pm$0.07 & $-2.8\pm 1.90$ \\
359.9309 & $-0.06064$ & 2.8 & Red & 1.9$\pm$0.09 & $-3.0\pm 1.72$ \\
359.9292 & $-0.05779$ & 2.7 & Red & 2.7$\pm$0.19 & $1.7\pm 1.46$ \\
359.9257 & $-0.05210$ & 2.8 & Red & 0.9$\pm$0.24 & $73.1\pm 8.68$ \\
359.9240 & $-0.04925$ & 2.9 & Red & 0.7$\pm$0.10 & $68.4\pm 4.30$ \\
359.9222 & $-0.04641$ & 3.1 & Red & 0.8$\pm$0.08 & $84.1\pm 3.74$ \\
359.9205 & $-0.04356$ & 3.4 & Red & 0.9$\pm$0.09 & $-85.7\pm 3.57$ \\
359.9407 & $-0.07028$ & 3.5 & Red & 1.3$\pm$0.12 & $61.4\pm 2.40$ \\
359.9372 & $-0.06459$ & 2.8 & Red & 2.9$\pm$0.14 & $12.5\pm 1.48$ \\
359.9355 & $-0.06174$ & 2.5 & Red & 1.8$\pm$0.12 & $-3.9\pm 1.36$ \\
359.9338 & $-0.05890$ & 2.4 & Red & 1.7$\pm$0.10 & $5.6\pm 2.21$ \\
359.9320 & $-0.05605$ & 2.2 & Red & 1.7$\pm$0.12 & $21.7\pm 2.10$ \\
359.9303 & $-0.05321$ & 2.2 & Cyan & 1.6$\pm$0.10 & $45.9\pm 1.77$ \\
359.9286 & $-0.05036$ & 2.3 & Cyan & 0.9$\pm$0.09 & $74.3\pm 2.35$ \\
359.9268 & $-0.04752$ & 2.5 & Cyan & 0.7$\pm$0.08 & $79.3\pm 3.03$ \\
359.9251 & $-0.04467$ & 2.7 & Cyan & 0.5$\pm$0.11 & $-68.1\pm 6.23$ \\
359.9233 & $-0.04183$ & 3.0 & Red & 0.7$\pm$0.09 & $-63.6\pm 3.93$ \\
359.9216 & $-0.03898$ & 3.4 & Red & 1.2$\pm$0.10 & $-62.0\pm 2.08$ \\
359.9199 & $-0.03613$ & 3.8 & Red & 1.6$\pm$0.17 & $-54.5\pm 2.83$ \\
359.9436 & $-0.06854$ & 3.2 & Red & 0.7$\pm$0.11 & $43.2\pm 4.06$ \\
359.9384 & $-0.06001$ & 2.1 & Red & 2.3$\pm$0.18 & $-13.5\pm 2.04$ \\
359.9366 & $-0.05716$ & 1.9 & Red & 1.0$\pm$0.09 & $8.7\pm 2.12$ \\
359.9349 & $-0.05432$ & 1.8 & Cyan & 0.8$\pm$0.06 & $30.5\pm 2.29$ \\
359.9331 & $-0.05147$ & 1.8 & Cyan & 1.0$\pm$0.05 & $46.3\pm 1.58$ \\
359.9314 & $-0.04863$ & 1.9 & Cyan & 0.9$\pm$0.03 & $73.2\pm 1.28$ \\
359.9297 & $-0.04578$ & 2.1 & Cyan & 0.9$\pm$0.07 & $89.4\pm 2.09$ \\
359.9279 & $-0.04293$ & 2.4 & Cyan & 1.0$\pm$0.09 & $-56.5\pm 2.73$ \\
359.9262 & $-0.04009$ & 2.7 & Cyan & 0.8$\pm$0.08 & $-51.9\pm 3.03$ \\
359.9245 & $-0.03724$ & 3.1 & Red & 1.3$\pm$0.12 & $-58.0\pm 2.68$ \\
359.9412 & $-0.05827$ & 1.8 & Red & 2.2$\pm$0.21 & $-1.2\pm 3.09$ \\
359.9395 & $-0.05542$ & 1.5 & Red & 0.4$\pm$0.07 & $0.5\pm 5.19$ \\
359.9377 & $-0.05258$ & 1.3 & Green & 0.6$\pm$0.07 & $45.2\pm 2.87$ \\
359.9360 & $-0.04973$ & 1.3 & Green & 1.0$\pm$0.05 & $49.8\pm 1.38$ \\
359.9343 & $-0.04689$ & 1.4 & Green & 0.8$\pm$0.03 & $73.9\pm 1.03$ \\
359.9325 & $-0.04404$ & 1.7 & Cyan & 0.8$\pm$0.06 & $-81.9\pm 1.65$ \\
359.9308 & $-0.04120$ & 2.0 & Cyan & 1.2$\pm$0.09 & $-41.6\pm 2.99$ \\
359.9290 & $-0.03835$ & 2.4 & Red & 1.7$\pm$0.21 & $-31.0\pm 3.57$ \\
359.9423 & $-0.05369$ & 1.1 & Green & 0.5$\pm$0.11 & $-22.0\pm 6.32$ \\
359.9406 & $-0.05084$ & 0.8 & Green & 0.2$\pm$0.12 & $27.2\pm 17.16$ \\
359.9388 & $-0.04800$ & 0.8 & Green & 0.8$\pm$0.08 & $61.1\pm 2.67$ \\
359.9371 & $-0.04515$ & 1.0 & Green & 0.7$\pm$0.03 & $77.1\pm 1.18$ \\
359.9354 & $-0.04231$ & 1.4 & Green & 0.5$\pm$0.05 & $-82.5\pm 2.64$ \\
359.9336 & $-0.03946$ & 1.8 & Cyan & 0.8$\pm$0.18 & $-33.1\pm 6.35$ \\
359.9469 & $-0.05480$ & 1.3 & Green & 1.1$\pm$0.16 & $-35.1\pm 3.49$ \\
359.9451 & $-0.05195$ & 0.8 & Green & 1.2$\pm$0.07 & $-34.0\pm 1.74$ \\
359.9434 & $-0.04911$ & 0.4 & Magenta & 3.8$\pm$0.12 & $-19.5\pm 1.01$ \\
359.9417 & $-0.04626$ & 0.4 & Magenta & 2.4$\pm$0.09 & $-17.5\pm 1.18$ \\
359.9399 & $-0.04342$ & 0.7 & Green & 0.4$\pm$0.04 & $-88.1\pm 3.17$ \\
359.9382 & $-0.04057$ & 1.2 & Green & 0.5$\pm$0.07 & $-71.0\pm 3.65$ \\
359.9365 & $-0.03772$ & 1.6 & Cyan & 1.1$\pm$0.26 & $-63.3\pm 7.31$ \\
359.9347 & $-0.03488$ & 2.1 & Red & 0.8$\pm$0.29 & $-78.7\pm 8.38$ \\
359.9602 & $-0.07013$ & 4.1 & Red & 0.2$\pm$0.18 & $54.5\pm 21.08$ \\
359.9532 & $-0.05875$ & 2.2 & Red & 0.4$\pm$0.12 & $-14.5\pm 8.60$ \\
359.9515 & $-0.05591$ & 1.7 & Red & 0.2$\pm$0.12 & $-69.7\pm 12.93$ \\
359.9497 & $-0.05306$ & 1.3 & Green & 0.8$\pm$0.08 & $-60.0\pm 3.48$ \\
359.9480 & $-0.05021$ & 0.8 & Green & 1.2$\pm$0.05 & $-38.3\pm 1.41$ \\
359.9463 & $-0.04737$ & 0.3 & Magenta & 4.2$\pm$0.15 & $-18.9\pm 1.05$ \\
359.9445 & $-0.04452$ & 0.2 & Magenta & 2.8$\pm$0.10 & $-17.7\pm 1.05$ \\
359.9428 & $-0.04168$ & 0.7 & Green & 0.2$\pm$0.04 & $-55.0\pm 5.59$ \\
359.9410 & $-0.03883$ & 1.1 & Green & 0.8$\pm$0.10 & $-51.7\pm 3.80$ \\
359.9630 & $-0.06840$ & 4.2 & Red & 0.2$\pm$0.12 & $12.2\pm 16.97$ \\
359.9613 & $-0.06555$ & 3.7 & Red & 0.8$\pm$0.16 & $14.9\pm 4.95$ \\
359.9560 & $-0.05701$ & 2.3 & Red & 0.9$\pm$0.18 & $-53.0\pm 5.08$ \\
359.9543 & $-0.05417$ & 1.8 & Red & 1.0$\pm$0.10 & $-67.9\pm 2.85$ \\
359.9526 & $-0.05132$ & 1.4 & Green & 1.2$\pm$0.07 & $-78.7\pm 1.77$ \\
359.9508 & $-0.04848$ & 1.0 & Green & 1.1$\pm$0.06 & $85.3\pm 1.68$ \\
359.9491 & $-0.04563$ & 0.7 & Green & 0.4$\pm$0.07 & $-83.7\pm 3.74$ \\
359.9474 & $-0.04279$ & 0.7 & Green & 0.2$\pm$0.07 & $-69.3\pm 10.08$ \\
359.9456 & $-0.03994$ & 0.9 & Green & 0.4$\pm$0.08 & $-71.2\pm 5.40$ \\
359.9439 & $-0.03710$ & 1.3 & Green & 0.2$\pm$0.10 & $83.8\pm 13.13$ \\
359.9422 & $-0.03425$ & 1.7 & Red & 1.0$\pm$0.19 & $-48.3\pm 5.36$ \\
359.9404 & $-0.03141$ & 2.2 & Red & 0.2$\pm$0.13 & $38.8\pm 18.73$ \\
359.9387 & $-0.02856$ & 2.6 & Red & 0.2$\pm$0.18 & $11.0\pm 17.85$ \\
359.9676 & $-0.06951$ & 4.7 & Red & 0.4$\pm$0.12 & $52.3\pm 6.97$ \\
359.9658 & $-0.06666$ & 4.2 & Red & 0.4$\pm$0.14 & $-22.5\pm 9.40$ \\
359.9572 & $-0.05243$ & 2.0 & Cyan & 1.1$\pm$0.08 & $-80.0\pm 2.13$ \\
359.9554 & $-0.04959$ & 1.7 & Cyan & 1.3$\pm$0.06 & $88.2\pm 0.85$ \\
359.9537 & $-0.04674$ & 1.3 & Green & 1.8$\pm$0.06 & $80.9\pm 0.92$ \\
359.9520 & $-0.04390$ & 1.1 & Green & 1.8$\pm$0.07 & $78.4\pm 1.16$ \\
359.9502 & $-0.04105$ & 1.1 & Green & 1.0$\pm$0.10 & $76.6\pm 2.39$ \\
359.9485 & $-0.03821$ & 1.3 & Green & 0.6$\pm$0.11 & $-87.8\pm 4.39$ \\
359.9467 & $-0.03536$ & 1.6 & Cyan & 0.3$\pm$0.10 & $53.1\pm 7.12$ \\
359.9450 & $-0.03252$ & 2.0 & Red & 0.4$\pm$0.10 & $3.0\pm 7.92$ \\
359.9433 & $-0.02967$ & 2.4 & Red & 0.6$\pm$0.09 & $10.1\pm 5.49$ \\
359.9415 & $-0.02682$ & 2.8 & Red & 1.1$\pm$0.15 & $-0.1\pm 4.11$ \\
359.9687 & $-0.06492$ & 4.4 & Red & 0.4$\pm$0.11 & $3.4\pm 8.31$ \\
359.9600 & $-0.05070$ & 2.3 & Red & 1.1$\pm$0.08 & $89.1\pm 2.91$ \\
359.9583 & $-0.04785$ & 2.0 & Cyan & 1.3$\pm$0.06 & $79.0\pm 1.39$ \\
359.9565 & $-0.04501$ & 1.8 & Cyan & 1.8$\pm$0.06 & $75.2\pm 0.82$ \\
359.9548 & $-0.04216$ & 1.6 & Cyan & 2.2$\pm$0.07 & $77.7\pm 1.02$ \\
359.9531 & $-0.03932$ & 1.6 & Cyan & 1.3$\pm$0.09 & $86.9\pm 1.98$ \\
359.9513 & $-0.03647$ & 1.7 & Red & 0.9$\pm$0.09 & $79.6\pm 3.20$ \\
359.9496 & $-0.03363$ & 1.9 & Red & 0.9$\pm$0.09 & $55.9\pm 3.34$ \\
359.9478 & $-0.03078$ & 2.3 & Red & 0.7$\pm$0.12 & $32.5\pm 4.90$ \\
359.9461 & $-0.02793$ & 2.6 & Red & 0.8$\pm$0.17 & $19.2\pm 6.03$ \\
359.9628 & $-0.04896$ & 2.7 & Red & 1.1$\pm$0.14 & $85.0\pm 3.70$ \\
359.9611 & $-0.04612$ & 2.4 & Red & 1.4$\pm$0.07 & $73.4\pm 1.42$ \\
359.9594 & $-0.04327$ & 2.2 & Cyan & 2.1$\pm$0.07 & $69.6\pm 1.03$ \\
359.9576 & $-0.04043$ & 2.1 & Cyan & 2.4$\pm$0.09 & $70.4\pm 1.09$ \\
359.9559 & $-0.03758$ & 2.1 & Cyan & 1.5$\pm$0.11 & $76.5\pm 1.72$ \\
359.9542 & $-0.03473$ & 2.2 & Red & 1.6$\pm$0.14 & $82.2\pm 2.26$ \\
359.9524 & $-0.03189$ & 2.3 & Red & 0.7$\pm$0.16 & $58.2\pm 7.74$ \\
359.9507 & $-0.02904$ & 2.6 & Red & 0.9$\pm$0.17 & $19.6\pm 5.17$ \\
359.9490 & $-0.02620$ & 2.9 & Red & 0.2$\pm$0.21 & $19.4\pm 18.47$ \\
359.9640 & $-0.04438$ & 2.8 & Red & 1.5$\pm$0.14 & $74.1\pm 2.72$ \\
359.9622 & $-0.04153$ & 2.6 & Red & 1.5$\pm$0.13 & $69.8\pm 2.28$ \\
359.9605 & $-0.03869$ & 2.6 & Cyan & 1.7$\pm$0.11 & $66.4\pm 1.87$ \\
359.9587 & $-0.03584$ & 2.5 & Red & 1.0$\pm$0.10 & $74.5\pm 3.05$ \\
359.9570 & $-0.03300$ & 2.6 & Red & 0.3$\pm$0.12 & $84.2\pm 12.11$ \\
359.9553 & $-0.03015$ & 2.8 & Red & 0.3$\pm$0.17 & $-17.8\pm 13.56$ \\
359.9633 & $-0.03695$ & 3.0 & Red & 0.9$\pm$0.18 & $74.3\pm 6.13$ \\
359.9616 & $-0.03411$ & 3.0 & Red & 0.4$\pm$0.16 & $16.8\pm 9.59$ \\
359.9599 & $-0.03126$ & 3.1 & Red & 1.0$\pm$0.22 & $9.1\pm 4.56$ \\
\end{longtable}

\clearpage
\begin{longtable}[!t]{*{6}{l}}
\caption{
Physical Properties of the Five Subregions in Figure \ref{fig:CS21spGuideMap} ---
$\Phi$ denotes the azimuthal-angle ranges of each subregion. 
Note that these ranges define the subregions, as described in \S\ref{sss:VF}.
$V_{\mathrm{LSR}}$ indicates the LSR-velocity range of the gas associated with each subregion (see the arrows on both sides of each spectrum in Figure \ref{fig:CS21spGuideMap}). 
The velocity ranges are identified by the ring-model shown in the central-upper panel of Figure \ref{fig:CS21spGuideMap} (see \S\ref{sss:VF} for the details). $\int I_{\nu}{\rm d}v$ represents the integrated intensity of the CS $J=2-1$ emission over the corresponding velocity range.
}\label{tbl:Regions}
\hline
Subregion &  A &  B &  C &  D &  E \\
\endfirsthead
\hline
\hline
\endhead
\hline
\endfoot
\hline
 $\Phi$ (degrees) & [$300, 350$] & [$200, 310$] & [$110, 200$] & [$13, 100]$ & [$10, 41$] \\
Area (pc$^2$)  & 1.1 & 1.3 & 0.71 & 1.0 & 1.2 \\
$V_{\mathrm{LSR}}$ (km\,s$^{-1}$) & [$-28, 110$] & [$-130, 58$] & [$-130, 40$] & [$-14, 120$] & [$62, 120$]\\
$\int I_{\nu}{\rm d}v$ \:(Jy $\cdot \mathrm{\:beam^{-1}\:km\:s^{-1}})$ & $3.9 \pm 0.07$ & $3.6 \pm 0.17$  & $4.3 \pm 0.10$ & $3.5 \pm 0.030$ & $2.2 \pm 0.10$\\
\end{longtable}

\begin{table}[h]
\caption{
Summary of the Model Fits to the 51 Segments within the 20$\sigma$-level Contour of the 37 $\mu$m Continuum Emission --- The terms Min, MAD$_{\rm stat}$, and Max denote the minimum, median absolute deviation, and maximum of the residuals, respectively. Here, MAD$_{\rm stat}$ is defined as $ median \left( \left| X_i - median(X) \right| \right)$. The errors of the MAD were estimated by approximating them with the standard error of the median.
The weighted-mean $|B_0^{\rm 3D}|$ in the last row was calculated using the $\chi^2$ values for the gc1, gc2, and gc3 models. 
This value is utilized in Eqs.~(\ref{eqn:alpha_WK}) and (\ref{eqn:beta_WK}) for the inverse calculation of \(\alpha_{\rm WK}\) and \(\beta_{\rm WK}\) values.
}
\label{tbl:model_fits}
\centering
\begin{tabular}{cccccclcc}
\hline
 &  & \multicolumn{3}{c}{Residual} & & \multicolumn{3}{c}{Estimated in \S\ref{sss:B0}} \\
 \cline{3-5} \cline{7-9}
\raisebox{1.5ex}[0pt][0pt]{Model} & \raisebox{1.5ex}[0pt][0pt]{$\chi^2$} & Min & $MAD_{\rm stat}$ & Max & & \raisebox{-1.5ex}[0pt][0pt]{$\alpha_{\rm WK}$} & \raisebox{-1.5ex}[0pt][0pt]{$\beta_{\rm WK}$} & $|B_0^{\rm 3D}|$ \\
 &  & (degrees) & (degrees) & (degrees) & & & & (mG) \\
\hline\hline
  gc1 & 29.8 & 0.59 & 6.0$\pm$3.2 & 88.3 & & 45 & 24 & 0.20 \\
  gc2 & 32.9 & 0.81 & 10.9$\pm$3.2 & 83.5 & & 35 & 19 & 0.23 \\
  gc3 & 33.4 & 0.44 & 13.4$\pm$3.1 & 88.4 & & 21 & 11 & 0.29 \\
  gc4 & 44.2 & 9.32 & 13.2$\pm$2.7 & 85.6 & & 22 & 12 & 0.29 \\
  gc5 & 60.3 & 14.49 & 15.1$\pm$2.8 & 89.1 & & 14 & 8 & 0.35 \\
\hline
 gc1, 2, 3 & $\cdot\cdot\cdot$ & $\cdot\cdot\cdot$ & $\cdot\cdot\cdot$ & $\cdot\cdot\cdot$ & & $\cdot\cdot\cdot$ & $\cdot\cdot\cdot$ & $0.24^{+0.05}_{-0.04}$\\ 
\hline
\end{tabular}
\end{table}

\clearpage
%
%
\appendix 

\end{document}